\documentclass[a4paper,fleqn]{cas-dc}

\usepackage[numbers]{natbib}
\usepackage{amsmath}
\usepackage{amssymb}
\usepackage{graphicx}
\usepackage{booktabs}
\usepackage{siunitx}
\usepackage{placeins}
\usepackage{float}
\usepackage{stfloats}
\usepackage{caption}
\usepackage{physics}
\newcommand{\vect}[1]{\mathbf{#1}} 
\newcommand{\vxB}{\vect{v} \times \vect{B}}
\newcommand{\vxvxB}{\vect{v} \times (\vect{v} \times \vect{B})}

\begin{document}
\let\WriteBookmarks\relax

\renewcommand{\topfraction}{0.9}
\renewcommand{\bottomfraction}{0.9}
\renewcommand{\textfraction}{0.05}
\renewcommand{\floatpagefraction}{0.5}
\setcounter{topnumber}{4}
\setcounter{bottomnumber}{4}
\setcounter{totalnumber}{8}

\shorttitle{A Bayesian Method for Air-Shower Reconstruction}

\shortauthors{Karen Terveer et~al.}

\title [mode = title]{A Bayesian Method for Air-Shower Reconstruction using Information Field Theory}


\author[1]{Karen Terveer}[orcid=0009-0002-9594-0419]
\cormark[1]
\ead{karen.terveer@fau.de}

\author[1]{Sjoerd Bouma}[orcid=0000-0002-6959-2302]
\author[2,3]{Stijn Buitink}[orcid=0000-0002-6177-497X]
\author[2,3]{Arthur Corstanje}[orcid=0000-0001-5992-6228]
\author[2]{Vital De~Henau}[orcid=0009-0003-0337-3558]
\author[4,5]{Vincent Eberle}[orcid=0000-0002-5713-3475]
\author[4,5,6]{Torsten A. En{\ss}lin}[orcid=0000-0001-5246-1624]
\author[7,4]{Philipp Frank}[orcid=0000-0001-5610-3779]
\author[8,2]{Tim Huege}[orcid=0000-0002-2783-4772]
\author[1]{Philipp Laub}[orcid=0000-0001-5610-3779]
\author[3,9]{Katharine Mulrey}[orcid=0000-0001-8026-8020]
\author[1,10]{Anna Nelles}[orcid=0000-0002-1720-6350]
\author[8]{Simon Str\"ahnz}[orcid=0009-0002-0494-4327]
\author[11]{Satyendra Thoudam}[orcid=0000-0002-7066-3614]
\author[8]{Keito Watanabe}[orcid=0009-0003-0337-3558]

\cortext[cor1]{Corresponding author}


\affiliation[1]{organization={Erlangen Centre for Astroparticle Physics, Friedrich-Alexander-Universit\"at Erlangen-N\"urnberg}, 
    address={91058 Erlangen}, country={Germany}}

\affiliation[2]{organization={Inter-University Institute For High Energies (IIHE), Vrije Universiteit Brussel (VUB)}, 
    address={Pleinlaan 2}, postcode={1050 Brussels}, country={Belgium}}

\affiliation[3]{organization={Department of Astrophysics/IMAPP, Radboud University Nijmegen}, 
    address={P.O.~Box 9010}, postcode={6500 GL Nijmegen}, country={The Netherlands}}

\affiliation[4]{organization={Max-Planck Institut f\"ur Astrophysik}, 
    address={Karl-Schwarzschild-Str.~1}, postcode={85748 Garching}, country={Germany}}

\affiliation[5]{organization={Ludwig-Maximilians-Universit\"at M\"unchen (LMU)}, 
    address={Geschwister-Scholl-Platz~1}, postcode={80539 M\"unchen}, country={Germany}}

\affiliation[6]{organization={Deutsches Zentrum f\"ur Astrophysik}, 
    address={Postplatz~1}, postcode={02826 Görlitz}, country={Germany}}

\affiliation[7]{organization={Kavli Institute for Particle Astrophysics \& Cosmology (KIPAC), Stanford University}, 
    address={CA 94305, Stanford}, country={USA}}

\affiliation[8]{organization={Institut f\"ur Astroteilchenphysik, Karlsruhe Institute of Technology (KIT)}, 
    address={P.O.~Box 3640}, postcode={76021 Karlsruhe}, country={Germany}}

\affiliation[9]{organization={Nikhef}, 
    address={Science Park Amsterdam}, postcode={1098 XG Amsterdam}, country={The Netherlands}}

\affiliation[10]{organization={Deutsches Elektronen-Synchrotron DESY}, 
    address={Platanenallee~6}, postcode={15738 Zeuthen}, country={Germany}}

\affiliation[11]{organization={Department of Physics, Khalifa University of Science and Technology}, 
    address={P.O.~Box 127788, Abu Dhabi}, country={United Arab Emirates}}

\begin{abstract}
The radio detection of extensive air showers provides a powerful method for studying the origin of high-energy cosmic rays. The Low-Frequency Array (LOFAR) offers unprecedentedly detailed measurements of the radio emission footprint. However, fully exploiting this information requires advanced reconstruction techniques. In this paper, we introduce a novel framework for air shower reconstruction based on Bayesian inference and Information Field Theory (IFT). Our method is built on a fully differentiable forward model of the radio signal, which incorporates a physical emission parameterization and a precise wavefront model. Additionally, we augment this physical model with Gaussian processes to account for systematic uncertainties in both the signal fluence and arrival timing. By leveraging gradient information, our approach enables efficient (three orders of magnitude acceleration w.r.t.\ the legacy method) and robust inference of the underlying physical shower parameters, such as primary energy and the depth of shower maximum, $X_\text{max}$. This work provides not only point estimates but also a rigorous quantification of uncertainties. We achieve a resolution in $X_\text{max}$ of \SI{25}{\gram\per\centi\meter\squared} and a radiation energy resolution of \SI{12}{\percent} on simulations for LOFAR. 
\end{abstract}

\begin{keywords}
Air shower reconstruction \sep Radio detection \sep Bayesian inference \sep Information Field Theory \sep LOFAR
\end{keywords}

\maketitle


\section{Introduction}

The origin of high-energy cosmic rays remains one of the most significant open questions in astroparticle physics. As charged particles, their trajectories are deflected by Galactic and intergalactic magnetic fields, obscuring the path to their astronomical sources. With their flux decreasing according to a power-law with energy, a direct detection at highest energies is no longer feasible. Our primary means of studying them is through the Extensive Air Showers (EAS) they initiate upon entering Earth's atmosphere. The properties of an EAS, particularly the energy of the primary particle ($E$) and the atmospheric depth of the shower maximum ($X_\text{max}$), from which statistically the mass composition can be inferred, are critical observables for constraining acceleration models and studying the mass composition of the cosmic-ray flux \cite{Kampert:2012}. 

For decades, EAS have been observed using ground arrays of particle detectors and optical telescopes that detect atmospheric fluorescence or Cherenkov light. Although successful, these techniques have limitations; for example, fluorescence detection is limited to clear nights without moon, resulting in a duty cycle of 10-15\% \cite{PierreAuger:2015nim}. In the last two decades, the detection of EAS radio emission has emerged as a complementary and standalone technique \cite{Huege:2016}. This emission provides a calorimetric measurement of the shower's electromagnetic component, is not significantly absorbed in the atmosphere, and can be detected with a near-100\% duty cycle.  

Previous work has shown excellent agreement between radio and other methods like Fluorescence detection \cite{PierreAuger:2023lkx}, Cherenkov-light detection \cite{Tunka-Rex:2015zsa}, and particle detectors \cite{PierreAuger:2016vya}, delivering superior per-event precision \cite{Buitink:2016nkf}. It has also been shown that the obtainable energy-scale has comparably small uncertainties and is consistent between experiments \cite{AbdulHalim:2025xz, Mulrey:2020oqe, Tunka-Rex:2016nto}. However, a consistent reconstruction that uses all information available in the electric field (i.e.\ amplitude, polarization, phase) without relying on very costly per-event simulations has been lacking. 

We present a new approach to analyzing the datasets provided by modern radio arrays. We introduce a novel, end-to-end framework for air shower reconstruction that leverages the principles of Bayesian inference. We treat the reconstruction as a probabilistic inference problem, which allows for the consistent propagation of uncertainties and the incorporation of prior physical knowledge. To address the high dimensionality inherent in analyzing the full radio signal field, we employ the formalism of Information Field Theory (IFT) \cite{Ensslin:2009}. 

IFT has already been applied to various astrophysical problems, including reconstructions of large-scale cosmological structures \cite{jasche2010}, mapping of the Galactic interstellar medium \cite{edenhofer2024}, and reconstructing electric fields of radio signals from cosmic rays and neutrinos \cite{welling2021,strahnz2024}. Recent works have demonstrated the potential of IFT in air-shower reconstruction using parameterized forward modeling of the radio signal \cite{terveer2025, strahnz2025} and template-based synthesis of the full electric-field signal \cite{watanabe2025}.

We demonstrate that this method can infer key shower parameters - arrival direction, core position, energy, and $X_\text{max}$ - along with their full posterior probability distributions, all while treating the air shower as a single phenomenon rather than a collection of separate observables, thereby maximizing the scientific potential of radio-detection arrays. This method simultaneously utilizes timing and fluence data with shared parameters, distinguishing it from previous approaches that relied on fluence only. Additionally, we will show that this method can reconstruct a larger selection of events than previous methods, allowing a future analysis of the complete LOFAR1.0 data set for a mass composition study. The treatment of the air shower as a random field additionally allows us to recover not just point estimates of the shower parameters, but their full posterior distributions and thus uncertainties on all parameters consistent with the underlying prior model and the data.

We present the method, tested on simulations and applied to a first selection of EAS measured with LOFAR.

\subsection{Radio emission in extensive air showers}\label{sec:radioemission}

\begin{figure*}[t]
    \centering
    \includegraphics[width=0.9\linewidth]{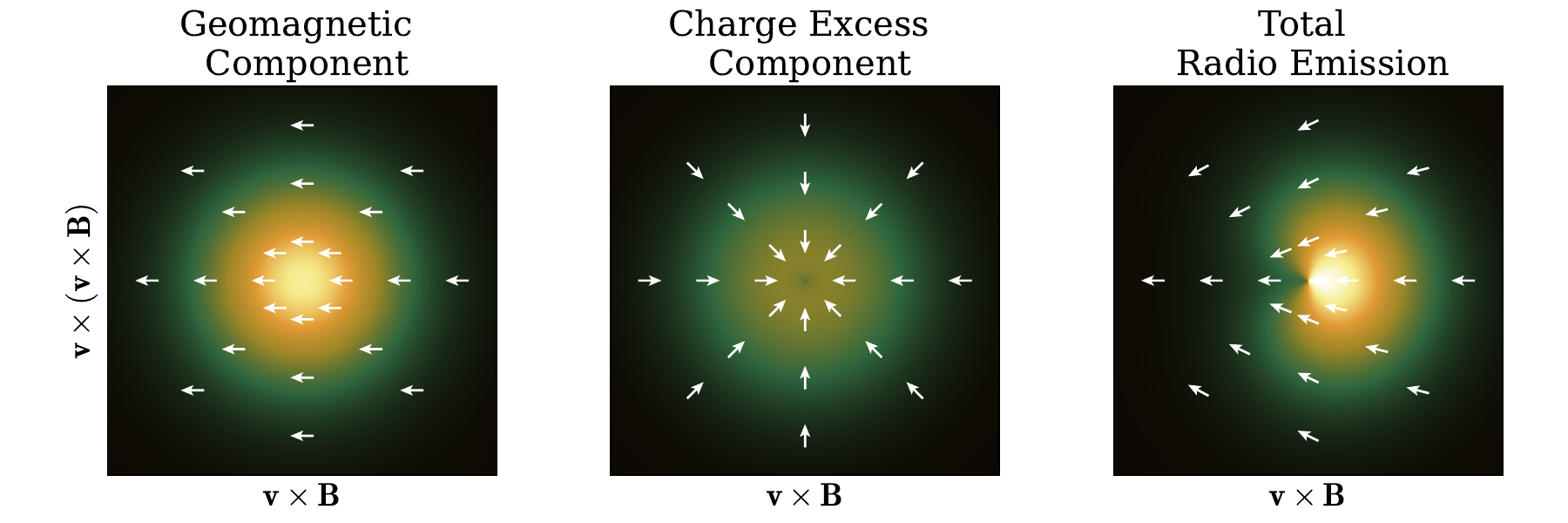}
    \caption{The radio pattern of the two emission mechanisms in the 30-\SI{80}{MHz} band, the Geomagnetic effect (left) and the Charge Excess effect (middle) along with the direction of their E-Field  indicated by arrows. The total emission (right) shows a characteristic asymmetric pattern due to the interference of the two effects. Patterns produced using parameterization from \cite{Glaser:2019}.}
    \label{fig:emissionmechanisms}
\end{figure*}

The charged particles in an EAS propagate in a thin disk called the \textit{shower front}, which propagates along the direction of the primary cosmic ray, the shower axis $\vect{v}$. There are two primary mechanisms producing radio emission in an air shower. The dominant one, also called the geomagnetic effect, arises due to the deflection of the electrons and positrons into opposite directions by the earth's magnetic field $\vect{B}$ in combination with their deceleration through interactions with the atoms of the atmosphere. The Lorentz force is given by $\vect{F}=q(\vect{v}\times\vect{B})$. The resulting charge separation creates a transverse current $\vect{J_\perp}$ flowing perpendicular to the shower axis. With the longitudinal development of the shower, this transverse current first rises and then dies out with the number of charged particles, causing a time-dependent change in the transverse current $\frac{\partial \vect{J_\perp}}{\partial t}$, which radiates electromagnetic waves \cite{Allan:1972wd,deVries:2011pa}. The strength of this emission scales with $\sin(\alpha)$, where $\alpha$ is the angle between the geomagnetic field and the shower axis, meaning that the geomagnetic effect is nearly zero if the shower axis is parallel to the geomagnetic field and the emission is strongest for showers perpendicular to it \cite{James:2010vm}. The emission from the geomagnetic effect is linearly polarized in the direction of the Lorentz force, $(\vect{v}\times\vect{B})$.

The second, generally sub-dominant effect is the charge excess effect, also referred to as Askaryan Effect \cite{Askaryan:1962}. As the shower develops, low-energy electrons are continuously stripped from atmospheric molecules via Compton scattering, and positrons from the shower annihilate with electrons. This results in the shower front accumulating a net negative charge excess, typically on the order of 10-20\% of the total number of particles \cite{Schellart:2014oaa}. As this net charge also varies in time while the shower develops, it radiates a sub-dominant radio pulse. The electric field from this effect is linearly polarized with the electric field vectors pointing radially inwards towards the shower axis in the shower plane.

As radio waves experience negligible attenuation in the atmosphere, the total emission from both effects reaches the ground. The two components superpose as $\vect{E}=\vect{E}_\text{geo}+\vect{E}_\text{ce}$, under the approximation that the components arrive in phase, yielding a purely linearly polarized signal. As both components have different polarization angles, constructive interference between the two on one side of the shower footprint generates a stronger signal, while the polarization vectors are opposite on the other side, weakening the electric field through destructive interference. Fig.~\ref{fig:emissionmechanisms} showcases this phenomenon and the resulting asymmetric radio footprint that reaches the ground.

As the emission is coherent at frequencies of tens of MHz, the electric fields from the individual particles add up constructively, resulting in the total radiated energy scaling with the number of particles in the shower ($E_\text{rad} \propto N^2$) and thus also scaling quadratically with the energy of the primary particle $E_\text{CR}$. The radio signal therefore provides a calorimetric measure of the electromagnetic component of the shower \cite{PierreAuger:2016vya}.

The emission is beamed forward because the shower front moves relativistically, and for precise calculations, it is important to account for the index of refraction of the atmosphere, which causes signal compression at the Cherenkov angle.
Consequently, the shower pattern is sensitive to the geometrical distance to the shower maximum and thus also the slant distance $D_{X_\text{max}}$ in \si{\gram\per\centi\meter\squared}.
To first order, this impacts the size of the footprint on the ground, as showers that developed at larger altitudes will produce much larger footprints than ones that develop closer to the ground \cite{Nelles:2014dja}.
The $D_{X_\text{max}}$ controls the shape of the radio footprint by determining how mature the air shower is when observed: showers close to the ground give a narrow, axis-centered peak, intermediate ones a broad Gaussian-like pattern, and far showers a distinct Cherenkov ring with a conical peak \cite{Glaser:2019}.

Sampling the radio footprint on the ground with an antenna array therefore provides information on the primary shower parameters. There are various radio arrays that observe data from EAS, like AERA \cite{PierreAuger:2025hoe}, and the Low-Frequency-Array (LOFAR) \cite{Schellart:2013bba}, which also serves as a pathfinder experiment for the next-generation radio array SKAO \cite{Corstanje:2025wbc}.

\subsection{Radio Detection of Air Showers with LOFAR}

The Low-Frequency Array (LOFAR) is a multi-purpose radio telescope with stations distributed across the Netherlands and several other European countries \cite{vanHaarlem:2013}. For cosmic-ray science, the key component is the dense core located in Exloo, the Netherlands. The core consists of a dense center region, the \textit{Superterp}, with six antenna fields within a \SI{320}{\meter} radius. With increasing distance there are 18 more core stations around the superterp. Each station contains 96 Low-Band Antennas (LBAs), operating in the 30 to \SI{80}{MHz} frequency range, and 48 high-band antennas (HBA, 110 to \SI{240}{MHz}). In the first operational stage of LOFAR, from 2012-2024, only the LBA were used for cosmic-ray measurements as the HBA data was beamformed and because LBA and HBA antennas could not observe simultaneously due to hardware limitations \cite{Schellart:2013bba,Nelles:2014dja}. With the upcoming LOFAR2.0 upgrade, the HBA data will be available together with the LBA data, enabling a sampling of the higher frequency footprint \cite{Mulrey:2025}. This current work, however, focuses on the 30-\SI{80}{MHz} band exclusively, but could in the future be extended to a broader frequency range.

For air shower detection, LOFAR operates in a triggered mode. An independent, co-located array of scintillator particle detectors, the LOFAR Radboud Air shower Array (LORA) \cite{Thoudam:2014}, provides an external trigger when an EAS is detected. Upon receiving a trigger, the time-series of raw Analog-To-Digital-Converter (ADC) counts from the LBA are saved for a short time window ($\sim$ \SI{2}{\milli\second}) around the event time. This captures the brief, nanosecond-scale radio pulse of the air shower. 

To convert ADC counts into physically meaningful units, an absolute calibration of the LOFAR antennas using a model of the Galaxy is used \cite{Mulrey:2019}, which yields a systematic uncertainty of \SI{13}{\percent}.

The standard method for reconstructing air shower properties with LOFAR \cite{Buitink:2014eqa} (here referred to as \emph{legacy}method ) is based on comparing fluence data to CoREAS \cite{Huege:2013} simulations. The fluence is a measure of energy deposit per area and can be calculated from the electric field via
\begin{equation}\label{fluence}
\mathcal{F}(\vect{r}_\mathrm{ant}) = \epsilon_0 c \sum_\mathrm{pol} \int_T \abs{E_\mathrm{pol} (t, \vect{r}_\text{ant})}^2\dd{t},
\end{equation}
where $E_\mathrm{pol}$ is the electric field trace in each E-field component, which varies as a function of time and antenna position.

In the legacy method, an analogous quantity to the fluence is calculated directly from voltage data of the instrumented measurement channels to prevent the passing of noise through the antenna model. This fluence-like quantity is then compared to simulated voltage data after passing the simulated electric fields through the forward response of the antenna.

For each measured shower, an ensemble of simulations is produced and their voltage-level fluences interpolated to LOFAR positions. These are then fitted to the measured pulse energies by minimizing a chi-squared function with free parameters for shower core position and an overall fluence scaling factor, and the resulting chi-squared values are plotted against $X_\text{max}$, with a parabolic fit to the lower envelope yielding the reconstructed $X_\text{max}$.

Uncertainties on $X_\text{max}$ are determined by adding realistic noise to the simulations and reconstructing them against the ensemble. The primary energy is estimated by applying the square root of the radio signal scaling factor to the simulation energy, given that the initial simulation energy is already close to the reconstructed energy. 

The mean uncertainty on $X_\text{max}$ of the method is \SI{19}{\gram\per\centi\meter\squared}, which is the current state-of-the-art \cite{Buitink:2014eqa}. The drawback of this method is its computational demand, as already one of the 60 CoREAS simulations that need to be produced per measured LOFAR event has a runtime of 1 CPU day. In the past year, an approach to decrease the computational load via template synthesis has become available \cite{DESMET2026}, but has not yet been used for a reconstruction study. 

Additionally, the legacy method relies solely on fluence data, discarding the information that other parts of the radio data contains. For example, the pulse arrival times also provide information on the $X_\text{max}$ \cite{Corstanje:2014waa,Apel:2014}. No established reconstruction approach has combined both timing and fluence so far. 

Our new approach relies on Bayesian inference, more specifically the Information Field Theory framework that is ideal for reconstructing high-dimensional, noisy data. It not only encompasses more information, but is also faster in reconstructing shower properties than the conventional analysis by relying on a semi-parametric air shower model instead of simulations, while retaining a higher accuracy than simple parametric approaches. 

\subsection{Information Field Theory}

Information Field Theory (IFT) provides a probabilistic framework for reconstructing physical fields by treating them as continuous random variables defined through probability distributions \cite{ensslin2019}. This Bayesian foundation allows the framework to address high-dimensional inference problems.

The objective of IFT is to infer an unknown signal field $\phi$ from a set of measurements $d$. Bayes' theorem expresses the posterior distribution as
\[
P(\phi|d) \propto P(d|\phi)\,P(\phi),
\]
where the likelihood $P(d|\phi)$ encodes the measurement process—including instrumental response and noise—while the prior $P(\phi)$ incorporates physical knowledge about the signal's statistical properties, such as its correlation structure.

In many cases, Gaussian statistics provide an adequate description:
\begin{align}
    P(d|\phi) &\propto \exp\Bigl(-\frac{1}{2}(d - R(\phi))^\dagger N^{-1} (d - R(\phi))\Bigr), \\
    P(\phi) &\propto \exp\Bigl(-\frac{1}{2}\phi^\dagger S^{-1} \phi\Bigr),
\end{align}
with $R(\phi)$ denoting the (possibly non-linear) response operator, $N$ the noise covariance matrix, and $S$ the signal covariance matrix. In its simplest incarnation, the inference problem reduces to minimizing the information Hamiltonian (which is equivalent to maximum a posteriori (MAP)), defined as the negative log-posterior:
\[
\mathcal{H}(\phi|d) = -\ln P(\phi|d).
\]

In variational inference (VI) approaches within IFT, the objective is to minimize the variational information distance between a variational distribution $Q(\phi)$ and the posterior $P(\phi|d)$, given by
\begin{align}
    D_\text{KL}(Q(\phi)||P(\phi))=-\mathbb{E}_Q\left[\ln\frac{P(\phi|d)}{Q(\phi)}\right]
\end{align}
where $D_{\text{KL}}$ is the Kullback-Leibler divergence and $\mathbb{E}_Q$ denotes the expectation with respect to $Q$. In its VI version (with $Q$ being the first argument), it measures the amount of spurious information introduced by approximating $P(\phi|d)$ by $Q(\phi)$. This should be as small as possible, hence the minimization \cite{knollmuller2020,frank2021}.

The VI approach of IFT enables the quantification of uncertainty by characterizing the posterior distribution rather than providing only point estimates, as given for example by MAP based methods. In non-linear settings, the posterior is typically non-Gaussian, preventing exact computation. To address this, approximation techniques based on VI have been developed within the IFT framework, such as Metric Gaussian Variational Inference (MGVI) \cite{knollmuller2020}. MGVI approximates the posterior with a Gaussian distribution that depends only on a posterior mean value for the quantities that are to be inferred, while the uncertainty covariance is expressed in terms of this mean with the help of the Fisher information metric. This improves computational efficiency compared to standard sampling methods. This efficiency is necessary for processing large datasets from radio observatories such as LOFAR and the SKA.

To address the limitations of the Gaussian approximation used by MGVI, Geometric Variational Inference (geoVI) \cite{frank2021} employs normalizing flows to learn a non-linear mapping from a simple base distribution to the target distribution. This enables the variational family to approximately represent non-Gaussian posteriors and represent intricate degeneracies, providing more detailed uncertainty quantification at the cost of increased computational demand.

In this work, we present a full air shower reconstruction method implemented using the Numerical Information Field Theory (NIFTy) framework in Python \cite{edenhofer2024niftyre,arras2019,steininger2019,selig2013}. NIFTy utilizes Google's JAX library \cite{bradbury2018} for automatic differentiation and just-in-time compilation, enabling GPU acceleration for efficient data processing.
\section{The Differentiable Forward Model}
Our method is centered around a Bayesian inference engine that connects a detailed physical forward model of the air shower signal to the observed data. A key innovation of our work is the development of a \textit{fully differentiable} forward model, which allows us to leverage powerful gradient-based inference techniques. 

The forward model, $F(\vec{\lambda})$, maps a set of physical air shower parameters, $\vec{\lambda}$, to the predicted data observables. It is constructed as a chain of modular, differentiable operations implemented in JAX.

\subsection{Coordinate Systems and Transformations}
The physics of the emission is most naturally described in the \textbf{shower-plane coordinate system}. This is a right-handed system with its origin at the shower core (i.e., impact) position on the ground, $(X_{\text{core}}, Y_{\text{core}})$, and its $\hat{z}'$-axis pointing upwards along the shower propagation direction $\vect{v}$. The transverse axes $(\hat{x}', \hat{y}')$ are aligned with the polarization vector of the geomagnetic ($\vxB$) emission and the direction perpendicular to it ($\vxvxB$). The parameterizations of our model are all described in this coordinate frame, however the forward model itself contains coordinate transformations such that its output is always in the ground plane ($\vect{x},\vect{y},\vect{z})$ (cartesian coordinates, with x pointing east) for a given arrival direction $(\theta, \phi)$ and core position. This transformation enables using ground plane data for the reconstruction, and embeds the shower-geometry-dependent coordinate transformation into the optimization, enhancing core and arrival direction precision and minimizing the geometry assumptions (i.e.\ known arrival direction) that need to be made pre-reconstruction.

\subsection{Fluence Model: The 'geoceLDF' Parametrization}\label{sec:ldfmodel}
We base our fluence prediction on the semi-analytical `geoceLDF` model \cite{Glaser:2019}, which provides a physically motivated Lateral Distribution Function (LDF) by combining the two primary emission mechanisms (Sec.~\ref{sec:radioemission}) in superposition. 
The fluence quantity we use is calculated from the electric fields in eV/m$^2$, not the voltage-level fluence previous analyses LOFAR used.

The parameterization is most accurate for showers with $\theta<\SI{60}{\degree}$, which is ideal for the dense LOFAR core. It models both the geomagnetic and the charge excess effect contributions in the 30 to \SI{80}{MHz} band separately, along with their dependence on the shower parameters $E_\text{rad}$, the slant distance $D_{X_\text{max}}$ and the shower direction $\theta, \phi$. Site-specific dependencies such as the magnetic field vector or the atmospheric model and observational height can be defined as well, making the model instrument-independent.

The total fluence at any position $\vect{r} = \left( \begin{matrix} r \\ \varphi_\text{obs} \end{matrix} \right)$ in the shower plane is a superposition that depends on the observer's polar angle relative to the shower core $\varphi_\text{obs}$: 
\begin{equation}
    \begin{split}
    \mathcal{F}(\vect{r}) = \mathcal{F}_{\text{geo}}(r) + \mathcal{F}_{\text{ce}}(r) \\
    + 2 \sqrt{\mathcal{F}_{\text{geo}}(r) \mathcal{F}_{\text{ce}}(r)} \cos(\varphi_\text{obs})
    \end{split}
\end{equation}

Through this dependence on observer angle, the interference caused by the different polarization patterns of the emission mechanisms can be caught.

We have implemented two critical enhancements to this model. First, we have extended it to directly accept the absolute atmospheric depth of the shower maximum, $X_\text{max}$ (in g/cm$^2$), as a primary input parameter. The $X_\text{max}$ and $D_{X_\text{max}}$ can, for less inclined showers ($\theta<$~\SI{60}{\degree}) be related via
\begin{equation}
    D_{X_{\text{max}}} = \frac{X_{\text{atm}}}{ \cos(\theta)} - X_{\text{max}},
\end{equation}
where $X_{\text{atm}}$ is the vertical atmospheric depth at ground level, which depends on the atmospheric model that can be chosen based on the site's atmospheric conditions. For real events, a unique atmospheric profile of the event time can be imported. This is a crucial modification, as $X_\text{max}$ is the key physical observable for mass composition studies, and an inaccurate atmosphere model can easily introduce biases of $\mathcal{O}(10\%)$ \cite{Mitra:2020mza}. 

Secondly, the entire model was re-implemented from the ground up to be end-to-end differentiable. This required a complete rewrite of the source code, but enables IFT's powerful gradient-based inference techniques.

\subsection{Arrival Time Model: Hyperbolic Wavefront}\label{sec:wavefrontmodel}

The temporal structure of the radio signal's arrival is modeled primarily using a hyperbolic wavefront, as measured by LOFAR \cite{Corstanje:2014waa}, which accounts for the refractive index of the atmosphere and the relativistic movement of the shower. The geometric arrival time, i.e. the time of the maximum of the Hilbert envelope of the total electric field, $\tau_{\text{geo}, i}$ at antenna $i$ is given by the parameterization derived for LOPES \cite{Huege2008, Apel:2014}:

\begin{equation}\label{eq:wavefront_parameterization}
    \begin{split}
    \tau_{\text{geo}, i} = \frac{1}{c} \Bigl( \sqrt{(d_i \sin\rho)^2 + (c \cdot b_{\text{offset}})^2} \\
    + z'_{i} \cos\rho \Bigr) + t_0.
    \end{split}
\end{equation}

\begin{figure}
    \centering
    \includegraphics[width=0.99\linewidth]{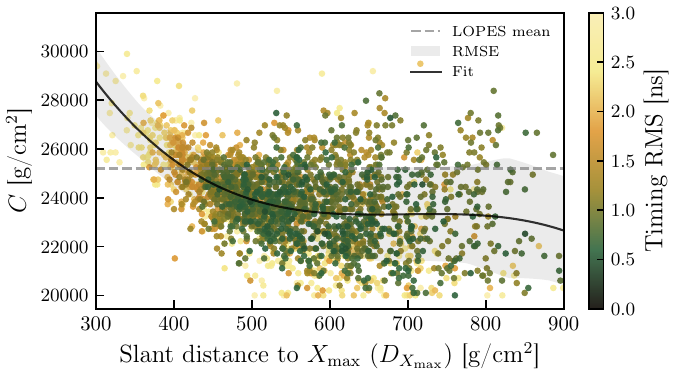}
    \caption{Dependence of the optimal timing parameter $C$ on the slant distance to the shower maximum, $D_{X_{\mathrm{max}}}$. Each point represents an individual CoREAS simulation, where the signal times were determined via the Hilbert envelope maximum of the total electric field trace, colored by the root-mean-square (RMS) of the timing residuals for the best fit. The solid black line shows the fit using a 3rd-degree polynomial (Tab. \ref{tab:poly_fit}), with the shaded region indicating the root-mean-square (RMS) deviations from the fit. The dashed gray line indicates the fixed value used by LOPES, $C = 25200~\mathrm{g/cm}^{2}$.}
    \label{fig:Cdxmax}
\end{figure}

Here, $d_i = \sqrt{x_i'^2 + y_i'^2}$ is the lateral distance from the shower axis, and $t_0$ is a global time offset. The parameter $\rho$ represents the cone opening angle, and $b_\text{offset}$ is fixed at \SI{3}{ns}. The $X_\text{max}$ dependence of the cone is described via 
\begin{equation}
    X_\text{max}=C\cdot \rho \cdot \cos ^{- \gamma}\theta,
\end{equation}
where $\gamma, C$ were fixed constants. We reparameterized these two constants on LOFAR simulations and found $\gamma=1.465 \pm 0.292$. For the constant $C$ we found a correlation with the slant distance to $X_\text{max}$, shown in Fig. \ref{fig:Cdxmax}, to which we fit a third degree polynomial (see Tab.~\ref{tab:poly_fit}).

\begin{table}
    \centering
    \renewcommand{\arraystretch}{1.3}
    \begin{tabular}{c|l}
         \textbf{Polynomial Term} & \textbf{Coefficient Value} \\ \hline
         $p_3 \cdot x^3$ & \SI{-8.955e-5}{(\gram\per\centi\meter\squared)^{-2}} \\
         $p_2 \cdot x^2$ & \SI{ 1.873e-1}{(\gram\per\centi\meter\squared)^{-1}} \\
         $p_1 \cdot x^1$ & \num{-1.302e2} \\
         $p_0$           & \SI{ 5.338e4}{\gram\per\centi\meter\squared}
    \end{tabular}
    \caption{Coefficients of the third-degree polynomial fit $C(x) = p_3 x^3 + p_2 x^2 + p_1 x + p_0$, where $x = D_{X_{\mathrm{max}}}$ in \unit{\gram\per\centi\meter\squared}.}
    \label{tab:poly_fit}
\end{table}

We note that the RMS in Fig.~\ref{fig:Cdxmax} indicates the goodness of fit decreases for smaller slant distances. This is likely due to the shower being closer to the ground. 

To account for potential tensions between the shower geometry preferred by the signal fluence and the wavefront timing we introduce a decoupling parameter, $\Delta X_{\text{max}}^{\text{time}}$. While the fluence model (Sec. \ref{sec:ldfmodel}) depends directly on the primary $X_{\text{max}}$, the wavefront parametrization utilizes an effective depth $X_{\text{max}}^{\text{time}} = X_{\text{max}} + \Delta X_{\text{max}}^{\text{time}}$. This nuisance parameter is constrained by a Gaussian prior with $\mu=0$ and $\sigma=\SI{15}{g/cm^2}$, allowing the inference to resolve minor inconsistencies between amplitude and timing data. This stabilizes the fit but we do not observe a significant bias introduced by it.

\subsection{Particle Model: NKG Function}\label{sec:particlemodel}

For completeness, our framework also includes a differentiable model for the ground-level particle density, based on the well-established Nishimura-Kamata-Greisen (NKG) function \cite{Greisen:1960, Kamata:1958}. This model predicts the lateral distribution of secondary particles as a function of shower parameters.

We note that the particle signal is significantly more complex to describe than the radio model due to stronger event-by-event fluctuations. In the current implementation, we find that including the particle data does not improve the reconstruction precision for key observables like $X_\text{max}$ or core position compared to a radio-only analysis, and causes tension. Consequently, the particle component is \textbf{not used in the primary reconstruction} presented in this work.

However, the model is retained for two important reasons: (1) it allows verification that the LORA particle detector array would have triggered on any given event, serving as an independent cross-check; and (2) it provides a foundation for future combined radio-particle analyses.

The full particle model is described in Appendix \ref{app:ParticleModel}.

\subsection{Modelling systematics with Correlated Fields}\label{sec:systematics}

Parametric models as described in Sections \ref{sec:ldfmodel} and \ref{sec:wavefrontmodel} are idealized representations of the complex statistical processes involved in an air shower. Based on simulations, they describe the idealized or at best average case of an air shower. In reality, the particle physics processes in an air shower undergo fluctuations, local atmospheric conditions \cite{Mitra:2020mza} or clouds \cite{Cantarini:2025} impact the propagation of radio signals,  and a perfect detector performance is never reached.

Standard least-squares fitting treats all deviations as uncorrelated white noise. However, many systematic effects are spatially correlated. By modeling correlation explicitly, we prevent the solver of the inference problem from treating systematic physical effects as random instrument noise.

Lastly, a purely parametric model is often too rigid (under-fitting), forcing physical parameters such as the shower maximum or core position to bias themselves to compensate for data that does not fit the ideal shape, leading to large $\chi ^2$-values and inaccurate posterior distributions. Adding correlated deviations to the model adds thousands of potential degrees of freedom (one for every point in space), but tames their behaviour via a prior that forces them to be a smooth function. This allows the model to bend where necessary to fit the data, without breaking the global parametric model.

IFT offers the modeling of correlations via Correlated Fields \cite{Arras:2022}. Correlated Fields do not treat the systematic effects as fixed parameters, but as continuous random fields with a Gaussian Process prior. In harmonic space, the power spectrum of the field we use is defined using a power law $P(k)\propto k^{-\alpha}$. A higher spectral index suppresses high-frequency modes, causing the field to be smoother and more correlated on large scales. A spectral index of 0 would correspond to uncorrelated white noise. 

To maintain differentiability in JAX, the field is not sampled directly, but modeled via a generative process.

By including these fields in the Bayesian inference, we effectively marginalize over the systematic uncertainties. The posterior distribution of the physical parameters (core, energy) widens, leading to more truthful and robust error estimates.

We introduce two Correlated Fields to our model, one acting on the fluence and one acting on the timing. As fluence variations scale with the signal strength, we apply the fluence Correlated Field multiplicatively, i.e. allowing it to alter each fluence point by a few percent. This is best achieved via an exponential factor $s_\text{syst}=\exp(s'_\text{syst})$, where the fluctuations of $s'_\text{syst}$ are modeled as normally distributed around zero. That way, if $s'_\text{syst}=0$ the Correlated Field becomes $=1$ and does not affect the fluence. Furthermore, we assign a spectral index of the field of $\alpha=-4.5 \pm 0.5$, favouring larger scale correlations.

The timing Correlated Field $\delta\tau_\text{syst}$ acts additively on the wavefront model. We define zero-centered, normally distributed fluctuations with $\sigma=\SI{1.5}{ns}$, and a spectral index of $\alpha=-1 \pm 0.5$, allowing smaller-scale variations.

Examples of posterior Correlated Fields are shown in Fig. \ref{fig:simreco} (third column from the left), where they are plotted exactly as they act on the parameterized models.

We apply nonlinearities (clipping) to the fields. This is critical when mixing physical and non-parametric models. It ensures the "flexible" field can correct local shapes but is mathematically forbidden from drifting so far that it replaces the parametric relations (e.g., the field cannot simply "create" the air shower signal from scratch, it can only modulate the physical fluence prediction by $\pm\SI{20}{\percent}$ and the timing prediction by $\pm \SI{15}{ns}$).

\subsection{The Full Model}
\label{sec:full_model}

Combining the physical parameterizations with the non-parametric systematic fields for both fluence and timing, the full forward model for the observables at antenna $i$ is:

\begin{align}
    \mathcal{F}_i(\vec{\lambda}, s_{\text{syst}}) &= \left[ \mathcal{F}_{i}(\vec{\lambda}_{\text{shower}}) \cdot \exp(s'_{\text{syst}, i}) \right], \\
    \tau_i(\vec{\lambda}, \delta \tau_{\text{syst}}) &= \tau_{\text{geo}, i}(\vec{\lambda}_{\text{time}}) + \delta \tau_{\text{syst}, i}.
\end{align}

The parameter vector that is to be inferred now includes the physical shower parameters $\vec{\lambda}$ and the latent variables defining both the fluence correlated field $s_{\text{syst}}$ and the timing correlated field $\delta \tau_{\text{syst}}$. Correlated fields can be turned "off" and "on" separately, also enabling a fully parametric forward model on demand.

\begin{figure*}[t]
    \centering
    \includegraphics[width=1.0\linewidth]{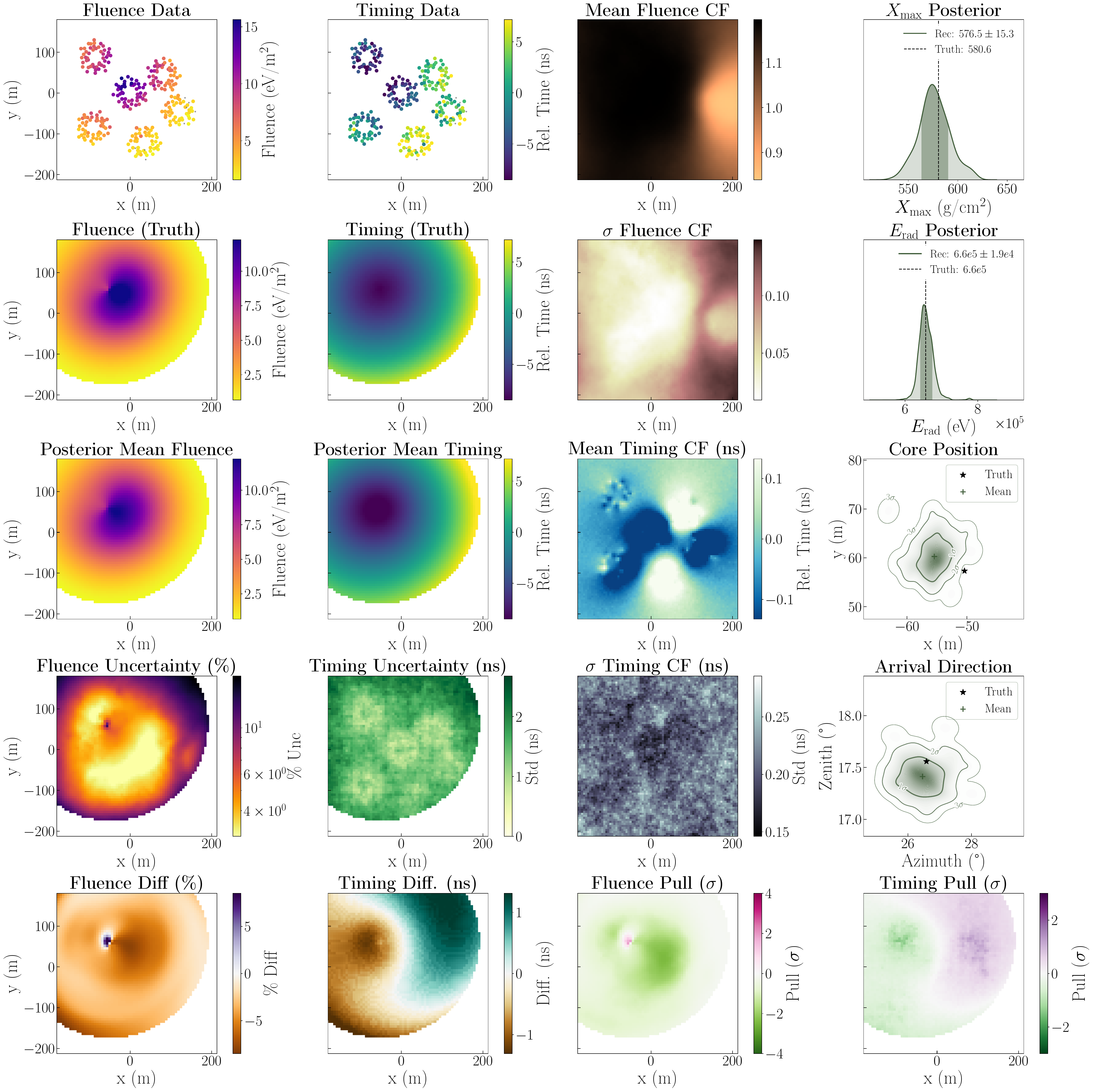} 
    \caption{Reconstruction summary for a simulated air shower event, an iron primary with an energy of \SI{2.44e+17}{eV}. The first two columns display the ground distributions of fluence and arrival timing (with the plane-wave component subtracted to better emphasize the hyperbolic shower front), showing (top to bottom) input data at LOFAR positions, CoREAS truth, reconstructed Posterior means, relative uncertainties, and residuals. The third column maps the inferred spatially Correlated Fields (CF). The fourth column presents the marginal posterior distributions for the depth of shower maximum $X_{\text{max }}$, radiation energy ($E_{\text{rad}}$), core position ($X_0, Y_0$), and arrival direction ($\theta, \phi$), with ground truth values indicated. The last row compares obtained differences with the corresponding pull.}
    \label{fig:simreco}
\end{figure*}

Figure \ref{fig:simreco} is an example reconstruction using this model. This visualizes the components involved in the forward model. For example, it shows the access to the parameter distributions ($X_\text{max}, E_\text{rad}$, arrival direction, core position) along with the spatial fluence and timing distributions. In principle, one can access the approximation of the full fluence or timing posterior distribution of each point in space, as well as the Correlated Fields at each point in space. This is difficult to visualize, so instead we show the mean and standard deviation of the posterior samples at each point.

The model is now semi-parametric. The fluence and timing parameterizations are deterministic, meaning with the same set of shower parameters the model will always return identical values. The Correlated Fields are what adds randomness to the model, allowing for small deviations given the same parameter set. 

The implementation of this entire chain—from the primary shower parameters, the parametric and non parametric components to the clipping operations - is done in the differentiable JAX framework. This allows us to calculate the analytical gradient of the posterior probability, $\nabla_{\vec{\theta}} \ln P(\vec{\theta}|d)$, unlocking efficient inference algorithms that can navigate this high-dimensional space without becoming trapped in local minima.

\section{Bayesian Inference and Parameter Estimation}\label{sec:likelihood}

Having established the generative forward model in Section~\ref{sec:full_model}, we now turn to the inverse problem: reconstructing the set of physical and systematic parameters, $\vec{\theta}$, given the observed radio measurements. We adopt a Bayesian framework to quantify the posterior probability distribution,
\begin{equation}
    P(\vec{\theta} \mid \boldsymbol{d}) \propto \mathcal{L}(\vec{\theta}) \, \pi(\vec{\theta}) ,
\end{equation}
where $\mathcal{L}(\vec{\theta}):=P(d|\theta)$ is the likelihood of the data and $\pi(\vec{\theta}):=P(\theta)$ represents the prior knowledge encoded in the model structure.

This section details the three main components of our inference strategy. First, we define the likelihood function, which accounts for the specific noise properties of the radio antennas and the masking of unreliable timing data. Second, we describe the numerical optimization scheme used to approximate the high-dimensional posterior via VI. Finally, we report on the computational cost and efficiency of this reconstruction pipeline compared to legacy simulation-matching approaches.

\subsection{The Likelihood and Noise Model}

The observed data vector for a single event consists of two distinct measurement types collected at $N_{\rm ant}$ antennas:
\begin{itemize}
    \item Fluence: $\mathcal{F}_i^{\rm obs}$ (in\,eV\,m$^{-2}$), $i=1,\dots,N_{\rm ant}$
    \item Arrival-time: $\tau_i^{\rm obs}$ (in seconds), $i=1,\dots,N_{\rm ant}$
\end{itemize}

Only a subset of antennas provide reliable timing information (after outlier rejection, see Sec. \ref{sec:pulse_finding}). We therefore introduce a binary mask vector $\boldsymbol{m} \in \{0,1\}^{N_{\rm ant}}$ such that $m_i = 1$ if antenna $i$ contributes a timing measurement and $m_i = 0$ otherwise.

The forward model presented in Sec.~\ref{sec:full_model} predicts, for any parameter configuration $\vec{\theta}$, the corresponding fluence and arrival time at every antenna position:
\begin{align}
    \boldsymbol{\mathcal{F}}^{\rm model}(\vec{\theta})
        &= \boldsymbol{\mathcal{F}}_{\rm phys}(\vec{\lambda}_{\rm shower})
           \,\odot\, \exp\bigl(\boldsymbol{s}'_{\rm syst}\bigr) ,\\[4pt]
    \boldsymbol{\tau}^{\rm model}(\vec{\theta})
        &= \boldsymbol{\tau}_{\rm geo}(\vec{\lambda}_{\rm time})
           + \boldsymbol{\delta\tau}_{\rm syst} ,
\end{align}
where $\odot$ denotes element-wise multiplication, $\boldsymbol{s}'_{\rm syst}$ is the interpolated and clipped fluence systematic field, and $\boldsymbol{\delta\tau}_{\rm syst}$ is the interpolated and clipped timing systematic field.

We model the measurement errors as independent Gaussian noises with \emph{known} (data-driven) variances. The full likelihood is therefore
\begin{equation}
    \mathcal{L}(\vec{\theta}) \equiv
    P\!\left( \boldsymbol{\mathcal{F}}^{\rm obs},\boldsymbol{\tau}^{\rm obs} \,\big|\, \vec{\theta} \right)
    = \mathcal{L}_{\mathcal{F}}(\vec{\theta})\;
      \mathcal{L}_{\tau}(\vec{\theta}) ,
    \label{eq:full_likelihood}
\end{equation}
with the two decoupled contributions
\begin{align}
    \mathcal{L}_{\mathcal{F}}(\vec{\theta}) &=
        \prod_{i=1}^{N_{\rm ant}}
        \mathcal{N}\!\bigl(\mathcal{F}_i^{\rm obs} \big|\;
            \mathcal{F}_i^{\rm model}(\vec{\theta}),\; \sigma_{\mathcal{F},i}^2 \bigr)
        \label{eq:lik_fluence}\\[6pt]
    \mathcal{L}_{\tau}(\vec{\theta}) &=
        \prod_{i=1}^{N_{\rm ant}}
        \Bigl[ \mathcal{N}\!\bigl(\tau_i^{\rm obs} \big|\;
            \tau_i^{\rm model}(\vec{\theta}),\; \sigma_{\tau,i}^2 \bigr)^{m_i} \nonumber \\
        &\quad \times \mathcal{N}\!\bigl(\tau_i^{\rm obs} \big|\;
            0,\; 10^{24}\,{\rm s}^2 \bigr)^{1-m_i} \Bigr] .
        \label{eq:lik_timing}
\end{align}

The second factor in Eq.~\eqref{eq:lik_timing} is a technical addition: for antennas that do \emph{not} contribute timing information ($m_i=0$), we impose an extremely weak constraint (variance $10^{24}\,{\rm s}^{2}$) centered on zero. This is numerically equivalent to ignoring those data points while still keeping the likelihood well-defined over the full $2N_{\rm ant}$-dimensional data space.

In matrix form the negative log-likelihood (up to an irrelevant constant) used by the IFT optimizer is
\begin{equation}
    -\!\ln\mathcal{L}(\vec{\theta}) \;=\;
    \frac{1}{2} \Bigl(\boldsymbol{d} - \boldsymbol{h}(\vec{\theta})\Bigr)^{\!\top}\!
    \boldsymbol{N}^{-1}\!
    \Bigl(\boldsymbol{d} - \boldsymbol{h}(\vec{\theta})\Bigr) ,
    \label{eq:neglogL}
\end{equation}
where the concatenated data and model vectors are
\begin{equation}
    \boldsymbol{d} \;=\;
    \begin{pmatrix}
        \boldsymbol{\mathcal{F}}^{\rm obs} \\[4pt]
        \boldsymbol{\tau}^{\rm obs}
    \end{pmatrix}\; \in \mathbb{R}^{2N_{\rm ant}},
    \quad
    \boldsymbol{h}(\vec{\theta}) \;=\;
    \begin{pmatrix}
        \boldsymbol{\mathcal{F}}^{\rm model}(\vec{\theta}) \\[4pt]
        \boldsymbol{\tau}^{\rm model}(\vec{\theta})
    \end{pmatrix}\!.
\end{equation}
The noise covariance matrix $\boldsymbol{N}$ is approximated as diagonal, which we find sufficient for our case as the off-diagonal elements are generally small \cite{Ravn2025}:
\begin{equation}
    \boldsymbol{N} = \operatorname{diag}\bigl(\sigma_{\mathcal{F},1}^2,\dots,\sigma_{\mathcal{F},N_{\rm ant}}^2, \sigma_{\tau,1}^2,\dots,\sigma_{\tau,N_{\rm ant}}^2\bigr),
    \label{eq:noise_cov_full}
\end{equation}
so that its inverse, the precision matrix, is simply
\begin{equation}
\label{eq:precision_matrix}
\begin{split}
    \boldsymbol{N}^{-1} &= \operatorname{diag}\bigl(1/\sigma_{\mathcal{F},1}^2,\dots,1/\sigma_{\mathcal{F},N_{\rm ant}}^2, \\
    &\quad 1/\sigma_{\tau,1}^2,\dots,1/\sigma_{\tau,N_{\rm ant}}^2\bigr).
\end{split}
\end{equation}

Because the noise covariance is diagonal and the forward model is implemented with JAX, which is equipped with autodifferentiation, the full gradient $\nabla_{\vec{\theta}}(-\ln\mathcal{L})$ is analytically exact and extremely cheap to evaluate, which is crucial for efficient sampling in the high-dimensional space spanned by the two correlated fields (approximately 180\,000 latent amplitudes in total for the $300\times300$ grid).

\subsection{Gradient-Based Variational Inference}
The posterior distribution $P(\vec{\theta}|d)$ is a high-dimensional function that cannot be computed analytically. We therefore use VI to find a tractable, approximate posterior distribution $Q(\vec{\theta})$. This is achieved by minimizing the Kullback-Leibler (KL) divergence between $Q$ and $P$.

We perform this optimization using the \texttt{NIFTy} library \cite{edenhofer2024niftyre,arras2019,steininger2019,selig2013}. We utilize NIFTy's \texttt{optimize\_kl} routine, which implements advanced VI methods such as geoVI \cite{frank2021}. To ensure robust convergence, our inference strategy employs a multi-stage approach. We begin with a resampling-based optimization (\texttt{nonlinear\_resample}) to explore the parameter space globally and avoid local minima, before switching to a gradient-based update mode (\texttt{nonlinear\_sample}) to refine the solution. The result is a multi-dimensional approximate posterior, $Q(\vec{\theta})$, whose mean provides the maximum a posteriori point estimates and whose covariance matrix provides a full quantification of parameter uncertainties and correlations.

\subsection{Computational Demand}

The reconstruction framework presented here offers a substantial improvement in computational efficiency compared to standard simulation-matching techniques.

A typical reconstruction requires an average wall-clock time of $48 \pm 13$\,minutes. The peak memory usage remains stable across events at approximately $6.2$\,GiB. Both values depend on the number of data points, and on whether correlated fields are activated or not. 
This efficiency is a direct consequence of the differentiability of the forward model, which allows the optimizer to navigate the high-dimensional parameter space using exact gradients rather than brute-force sampling.

In contrast, the legacy reconstruction method relies on $\sim 60$ individual simulations per measured event. With the generation of a single detailed simulation taking roughly 1 CPU day, the computational demand of the legacy reconstruction, based on generating the simulations alone, amounts to approximately 1440 CPU hours per event, not including the demand of the consequent fitting to the data.

Therefore, the method proposed in this work reduces the computational cost by three orders of magnitude—from weeks of CPU time to approximately 1 CPU hour per event. This renders the detailed reconstruction of large datasets feasible on standard computing clusters without requiring massive high-performance computing allocations, while retaining accuracy superior to simple parametric reconstructions. 
\section{Performance on simulations}\label{sec:simtest}

The first test of this framework is performed on CoREAS simulations. This is the best way to compare the reconstructed shower properties to the truth and fully understand the methods biases and limitations.

\subsection{Generation of realistic data}
The simulations used are a pre-existing set for LOFAR conditions, with primary energies between $10^{16.5}$\,\si{\electronvolt} and $10^{18.5}$\,\si{\electronvolt} and $X_\text{max}$ values between 500 and \SI{1000}{\gram\per\centi\meter\squared}. The set contains both iron and proton primaries and used the hadronic interaction model QGSJetII-04 \cite{Ostapchenko_2013}. We are limiting our dataset to near-vertical simulations with a zenith angle $< \SI{45}{\degree}$, consistent with the geometry of typical LOFAR events. Showers at higher inclinations produce larger radio footprints where the energy is distributed over a larger area, causing signals to drop below detection threshold. Additionally, only a small amount of particles reach sea-level at high zenith angles, resulting in LORA not triggering for these events, resulting in LOFAR datasets being dominated by lower zenith angles.

For convenience we use the NuRadioReco software \cite{GlaserNRR:2019,Bouma:2025dmo} to read in CoREAS simulations and interpolate their electric field traces to LOFAR positions using Fourier-based interpolation \cite{Corstanje:2023}, assigning a random core from $\mathcal{N}(0,100)$\,\si{m}, where \SI{0}{m} is the center of the LOFAR superterp. This core distribution was chosen as it matches the distribution of cores in the LOFAR mass composition analysis \cite{Corstanje:2021}.

We then converted the interpolated electric fields to voltages using the LOFAR antenna response.
In order to more realistically simulate the behaviour of real data with simulations, we added measured noise from LOFAR data at the voltage level to the traces. Previous analyses used the voltages directly to calculate a fluence-like quantity to avoid passing noise through the antenna response. However, as our forward model expects the fluence in units of \si{\electronvolt\per\meter\squared}, we need the electric-field fluence. We thus add measured voltage noise at this stage to simulate the passing of noise through the antenna response.

For all subsequent steps of the reconstruction, the true arrival direction and core position of the simulation is randomized slightly within the known LORA precision \cite{Thoudam:2014} ($\sigma_{\theta,\phi}=\SI{0.7}{\degree}$ and $\sigma_{X_C,Y_C}=\SI{6}{m}$ for cores within \SI{150}{m} of the LORA center, and a more generous $\sigma_{\theta,\phi}=\SI{2}{\degree}$ and $\sigma_{X_C,Y_C}=\SI{20}{m}$ for cores further out), to simulate that in real data we would not know the true geometry either. 

We want to focus only on events that would have realistically been detected by LOFAR. For previous analyses, only simulations where more than \SI{50}{\percent} of antennas in at least 3 stations had an energy signal-to-noise ratio ($\frac{\mathcal{F}_\text{signal}^{55~\text{ns}}}{\mathcal{F}_\text{noise}^{55~\text{ns}}}$) of $> 6$ were included \cite{Corstanje:2021}. This was a conservative filter to prevent composition bias, which we will loosen to require only 1 station to fulfill this check. We thus allow both smaller and fainter footprints in our analysis, widening the prospective real dataset. We note that nevertheless, for a future mass composition analysis using this method, further tests against composition bias will need to be implemented.

\subsection{Pulse finding}\label{sec:pulse_finding}
If an event is considered \textit{triggered} (i.e. at least one station has more than \SI{50}{\percent} of antennas with an energy SNR $> 6$), it enters the pulse finding stage. For this, the voltage traces are first converted into electric fields via 
\begin{equation}
\begin{pmatrix}
V_1(f) \\
V_2(f)
\end{pmatrix}
=
\begin{pmatrix}
\mathcal{H}_1^\theta(f) & \mathcal{H}_1^\phi(f) \\
\mathcal{H}_2^\theta(f) & \mathcal{H}_2^\phi(f)
\end{pmatrix}
\begin{pmatrix}
E^\theta(f) \\
E^\phi(f)
\end{pmatrix},
\end{equation}
where $\mathcal{H}$ is the antenna response, $V$ are the voltages from the two antenna polarizations, using the arrival direction guess and solving for the electric field components $E$. 

Finding the exact pulse location is essential for a successful reconstruction, however especially at low SNR, pulse finding proves challenging.

Before analyzing individual antennas, we perform a rubust "incoherent beamforming" sum of all antennas per station to find a global timestamp estimate ($t_\mathrm{BF}$), which will then define the region of interest (ROI). This is not performed globally as it assumes a plane wavefront, which suffices as approximation only for smaller subsections of the footprint.

For arrival direction $(\theta,\phi)$, the plane-wave arrival time delay $\tau_i$ for antenna $i$ at position $(x_i,y_i)$ relative to the station center $(x_0,y_0)$ is 
\begin{equation}
    \begin{split}
    \tau_i = \frac{1}{c} \Bigl( \sin\theta\cos\phi \cdot (x_i-x_0) \\
    + \sin\theta\sin\phi \cdot (y_i-y_c) \Bigr).
    \end{split}
\end{equation}
We calculate the magnitude of the sum of electric-field components of each trace and apply a Gaussian smoothing ($\sigma \approx \SI{10}{\nano\second}$) to create a broad pulse. This is to ensure that even if the arrival direction guess is slightly off, the pulses from different antennas still overlap when summed.

The smoothed envelopes are then shifted by their geometric delays and averaged:
\begin{equation}
    S_\text{beam}(t) = \frac{1}{N_\text{ant}}\sum_{i=1}^{N_\text{ant}}E_{i,\text{smooth}}(t-\tau_i)
\end{equation}

The time of the resulting maximum is then used as the center of the ROI for the pulse search in individual antennas. In this region of interest, the maximum of the Hilbert envelope of the trace is calculated (on the non-smoothed trace). This is the simplest way to find pulses, and is successful for most signals. 

Next, we perform a cleaning via a local outlier rejection. Each antenna is compared to its 10 nearest neighbors by calculating the median and the standard deviation of all their timing values. If the antenna timing deviates by $2.5\sigma$ from the median value, the timing value is given no weight in the reconstruction, as described in Sec. \ref{sec:likelihood}. This pre-processing of the timing data is necessary as any strong deviations from the true timing values will massively impact the reconstruction performance, which can happen in low-SNR signals. 

This is not the case for fluence data. For very low SNR, where the timing guess might be hundreds of nanoseconds off, the calculated fluence value will not differ much from its true counterpart, as noise dominates here. The low-signal antennas are important to include in the data, as they provide a valuable upper limit for the footprint extent. 

Thus, for all antennas in a triggered station, the fluence is calculated via Eq. \eqref{fluence} in a time window of \SI{300}{ns} around the determined pulse maximum. This ensures that the entire signal is well within the fluence calculation window, even if the determined pulse time is slightly off. A window of \SI{300}{ns} is larger than usually used in analyses and results in a larger noise fraction in the fluence, however we find the advantages of having more of the signal in the window outweigh the disadvantage of a larger noise fraction. A smaller signal window causes the fluence to be underestimated which propagates into an energy underestimation in the reconstruction. 

\subsubsection{Modeling of fluence noise $\sigma_{\mathcal{F},i}$}

We do not perform any noise subtraction but pass the raw, noisy fluence as data into the optimization. Accordingly, we add a log-normal noise scale prior $\sigma_\text{noise}$ to the fluence forward model, which is calculated as the mean fluence noise from multiple non-overlapping noise windows (see Tab. \ref{tab:priors}).

Furthermore, for every antenna $i$ the fluence uncertainty is taken to be the quadratic sum of two independent contributions:
\begin{equation}
    \sigma_{\mathcal{F},i}^2 \;=\;
    \sigma_{\mathcal{F},i}^{\rm stat\,2} \;+\;
    \sigma_{\mathcal{F},i}^{\rm syst\,2}
    \;=\;
    \sigma_{\mathcal{F},i}^{\rm stat\,2} \;+\;
    (0.1 \cdot \mathcal{F}_i^{\rm obs})^2 .
    \label{eq:fluence_error}
\end{equation}
The statistical term $\sigma_{\mathcal{F},i}^{\rm stat}$ is estimated from the standard deviation of multiple non-overlapping noise windows. The 10\,\% systematic uncertainty is added in quadrature and dominates at high signal-to-noise ratios. The systematic uncertainty arises from both the absolute calibration of the antennas and also the finite sampling in the time domain.

\subsubsection{Modeling of timing noise $\sigma_{\tau,i}$}\label{sec:timing_noise}
The timing noise is more complicated, as it cannot directly be calculated from data and depends on the signal strength of the event. For high SNR events, the time of the maximum signal will usually be within \SI{5}{ns} (sampling time of LOFAR) of the true signal time, but for lower SNR events, due to the noise, the maximum might jump back or forth a few samples. Using the same timing noise estimate for all data points is not an option, so we need to estimate the timing noise on a per-antenna basis using local clusters.

A global plane-wave fit is insufficient because its residuals combine systematic error from the wavefront curvature with the true random noise, which would incorrectly inflate the uncertainty estimate and reduce sensitivity to $X_{\rm max}$. To address this, we perform a polynomial fit per antenna that includes a radial curvature term, capturing both the plane-wave delay and the wavefront curvature. For each antenna $i$, we define a local cluster $C_i$ consisting of its $k=20$ nearest neighbors. Positions within this cluster are centered and normalized for numerical stability:
\begin{equation}
    \tilde{x}_j = \frac{x_j - \bar{x}_{C_i}}{\sigma_r}, \quad \tilde{y}_j = \frac{y_j - \bar{y}_{C_i}}{\sigma_r}, \quad \tilde{r}_j^2 = \tilde{x}_j^2 + \tilde{y}_j^2,
\end{equation}
where $\sigma_r$ is the standard deviation of radial distances within the cluster. A least-squares fit is performed using the design matrix $\mathbf{A} = [1, \tilde{x}, \tilde{y}, \tilde{r}^2]$, solving for coefficients $\vec{c}$ such that $t_j \approx \mathbf{A}_j \cdot \vec{c}$. The residuals $\epsilon_j = t_j - \mathbf{A}_j \cdot \vec{c}$ represent pure measurement noise, free from systematic curvature contributions. The per-antenna timing uncertainty is then:
\begin{equation}
    \sigma_{\tau,i} = \max\left( \text{std}(\{\epsilon_j\}_{j \in C_i}), \sigma_{\tau}^{\min} \right),
\end{equation}
where $\sigma_{\tau}^{\min} = \SI{1.5}{ns}$ is a floor value to prevent unrealistically small uncertainties. If a cluster has fewer than 8 antennas, a fallback uncertainty of \SI{5}{ns} is used.

For antennas that pass all quality cuts ($m_i=1$), this per-antenna uncertainty is used. For antennas that are excluded from the timing fit ($m_i=0$), the effective uncertainty is set to $\sigma_{\tau,i} = 10^{12}\,{\rm s}$ (see Eq.~\eqref{eq:lik_timing}).

\subsubsection{Estimation of the time offset $t_0$}
The global time offset parameter $t_0$ (Eq. \eqref{eq:wavefront_parameterization}), essentially the time at the shower core, requires a specific prior derivation. While the geometric parameters are centered on the LORA reconstruction, $t_0$ is derived from the radio timing data itself. We utilize the spatial polynomial fit described in Sec. \ref{sec:timing_noise} and extrapolate it to the estimated core position. The value of this fit at the core defines the mean of the $t_0$ prior ($\mu_{t_0}$), while the prior width ($\sigma_{t_0}$) is derived from the covariance of the fit combined with a distance-dependent uncertainty term to account for extrapolation errors.

This ensures that the $t_0$ prior does not prevent the optimizer from being able to fit the timing data by being too narrow, which is especially important for those events with cores outside of the LOFAR superterp, where the $t_0$ estimate is usually less accurate.

\subsubsection{Prior distributions}\label{sec:hyperpriors}

We apply constraints on the hyperparameter prior distributions as displayed in Tab.~\ref{tab:priors}.
The radiation energy $E_{\mathrm{rad}}$ is sampled in logarithmic space, while geometric parameters are constrained by a uniform prior centered around the initial reconstruction provided by the LORA particle detector array.

\begin{table}
    \centering
    \caption{Prior distributions for the scalar shower parameters. $\mathcal{N}(\mu, \sigma)$ denotes a Normal distribution, $\mathcal{LN}(\mu, \sigma)$ denotes a Log-Normal distribution, and $\mathcal{U}(a, b)$ denotes a Uniform distribution. The geometric priors are dynamically centered on the initial guess $(\phi_{\text{init}}, \theta_{\text{init}})$ and particle core $(\mathbf{x}_{c, \text{init}})$.}
    \label{tab:priors}
    \renewcommand{\arraystretch}{1.3}
    \begin{tabular}{l l l}
        \toprule
        \textbf{Parameter} & \textbf{Symbol} & \textbf{Prior Distribution} \\
        \midrule
        Radiation energy & $E_{\mathrm{rad}}$ & $\mathcal{LN}(\ln(10^7), 3.0)$ [eV] \\
        Shower Azimuth & $\phi$ & $\mathcal{U}(\phi_{\text{init}} - 3^{\circ}, \phi_{\text{init}} + 3^{\circ})$ \\
        Shower Zenith & $\theta$ & $\mathcal{U}(\theta_{\text{init}} - 3^{\circ}, \theta_{\text{init}} + 3^{\circ})$ \\
        Core Position (X) & $X_{c}$ & $\mathcal{U}(X_{c, \text{init}} - w, X_{c, \text{init}} + w)$\textsuperscript{\textdagger} \\
        Core Position (Y) & $Y_{c}$ & $\mathcal{U}(Y_{c, \text{init}} - w, Y_{c, \text{init}} + w)$\textsuperscript{\textdagger} \\
        Depth of Max. & $X_{\max}$ & $\mathcal{U}(450, 1020)$ [g/cm$^2$] \\
        Time Offset & $t_0$ & $\mathcal{N}(\mu_{t_0}, \sigma_{t_0})$ \\
        Noise Scale & $\sigma_{\text{noise}}$ & $\mathcal{LN}(\mu_{\text{data}}, 0.1 \cdot \mu_{\text{data}})$ \\
        \bottomrule
    \end{tabular}

    \footnotesize{\textsuperscript{\textdagger} The prior width for the core position $w$ is based on the core location $d_c$. Within \SI{150}{m} of the LORA center, we can assume the resolution of the core guess is very precise \cite{Thoudam:2014}, while outside this radius the resolution is worse. Hence, we use $w_{d_c \leq \SI{150}{m}} = \SI{30}{m}$ and $w_{d_c > \SI{150}{m}} = \SI{60}{m}$.}
    
\end{table}

On simulations, we found the best reconstruction results when deactivating the fluence Correlated Field (Sec.~\ref{sec:systematics}), as spatial deviations on the fluence can cause biases in $X_\text{max}$ by distorting the footprint shape. It might be beneficial to re-activate it for reconstructing data, and making the field more noise-like by lowering the fields spectral index. This could help with per-antenna calibration or hardware differences. As the impact on reconstruction bias is complex, this would need further investigation before using it, thus we only show the fluence Correlated Fields for example events, but do not use them for the simulation or data sets we reconstruct in this work.

We use a correlated field on timing data, with a spectral index of $\alpha=-1 \pm 0.5$, allowing spatial correlations over smaller distances, but capping the field at $\pm$\SI{10}{ns}. We notice (see Fig.~\ref{fig:simreco}) that the correlated field appears to vary by antenna/station positions, indicating that it "irons" out station and antenna-wise timing offsets.

\subsection{Reconstruction performance}
Post-optimization, the forward model can be evaluated for the generated samples to obtain Posterior distributions of both the shower parameters and the full forward model. This allows studying the results beyond just point estimates. An example reconstruction result is shown in Fig. \ref{fig:simreco}. 

For the purpose of this example plot, we activated the fluence Correlated Field to demonstrate the model's capacity to infer spatial systematics. The plot shows a good agreement between truth and reconstructed signal, with absolute fluence residuals of less than \SI{10}{\percent} and timing residuals within \SI{2}{ns}.

Of all 500 simulations in the set, 390 (\SI{78}{\percent}) passed the post-reconstruction quality cuts,
demanding the reduced $\chi^2$ to be $<1.25$ and the core reconstruction uncertainty to be below $\SI{10}{\meter}$. These cuts ensure a high reconstruction quality without removing too many events. These cuts will be applied to real data as well.

We find that especially for higher energy events, the posteriors resulting from the reconstruction are too narrow. For lower energies, noise dominates the data, which is reflected in broader posterior distributions. For higher energy events, systematic data-model differences dominate, resulting lower uncertainties. Activating the fluence correlated field could be a solution to that in the future. For now, we perform a calibration of uncertainties based on the reconstructed radiation energy. The full calibration procedure is detailed in Appendix \ref{app:uncertainty_calibration}. This calibration has been applied to all of the results that are discussed in the following sections.

We investigated, whether this underestimation of uncertainties is inherently due to the usage of VI, which in some cases can lead to too narrow posteriors. Running the reconstruction again using a No U-Turn Sampler (NUTS) from the NIFTY framework revealed no significant change in posterior width, indicating that the under-coverage stems from the model and not the inference engine.

\subsubsection{$E_\text{rad}$ reconstruction performance}\label{sec:e_rad_reco}
\begin{figure*}[t]
    \centering
    \includegraphics[width=0.9\linewidth]{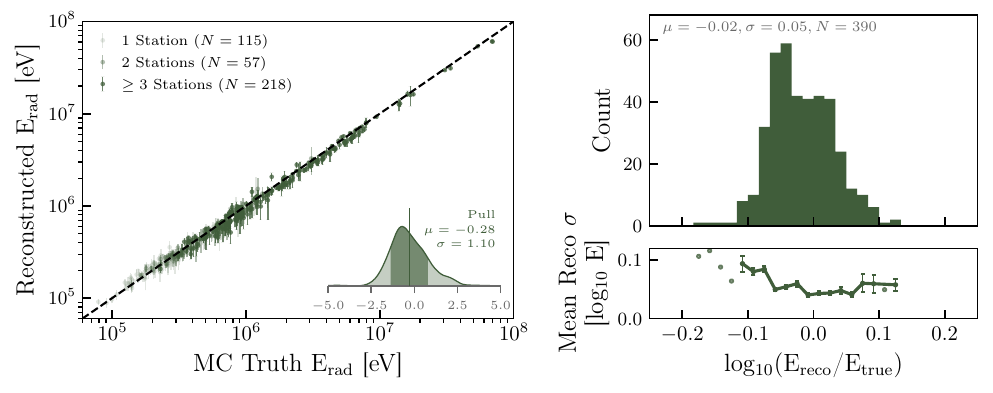}
    \caption{Performance of the radiation energy $E_\text{rad}$ reconstruction on CoREAS simulations. The left side displays both the reconstructed values compared to the CoREAS truth for all 390 successful reconstructions, as well as the pull from the reconstruction. The color-coding of the scatter points is based on how many stations were considered "triggered" for this specific event. The right side of the plot shows a histogram of the reconstruction errors at the top and how the reconstruction uncertainties correlate with reconstruction error.}
    \label{fig:erad_sim}
\end{figure*}

The performance of the reconstruction of the radiation energy $E_\text{rad}$ for the whole simulation set is shown in Fig.~\ref{fig:erad_sim}. 

We achieve an overall precision on $E_\text{rad}$ of \SI{11.5}{\percent} with a bias of \SI{4.5}{\percent}, leading to an overall resolution of \SI{12.4}{\percent}. The Figure also displays a Pull, which is an excellent measure of how well the reconstructed uncertainties are estimated. Post-calibration, the $\sigma$ of the Pull is 1.1. A pull $\sigma$ of 1 would correspond to ideal overall coverage. Pre-calibration the uncertainties were underestimated by an average factor of 4.45.

\subsubsection{$X_\text{max}$ reconstruction performance}
\begin{figure*}
    \centering
    \includegraphics[width=0.9\linewidth]{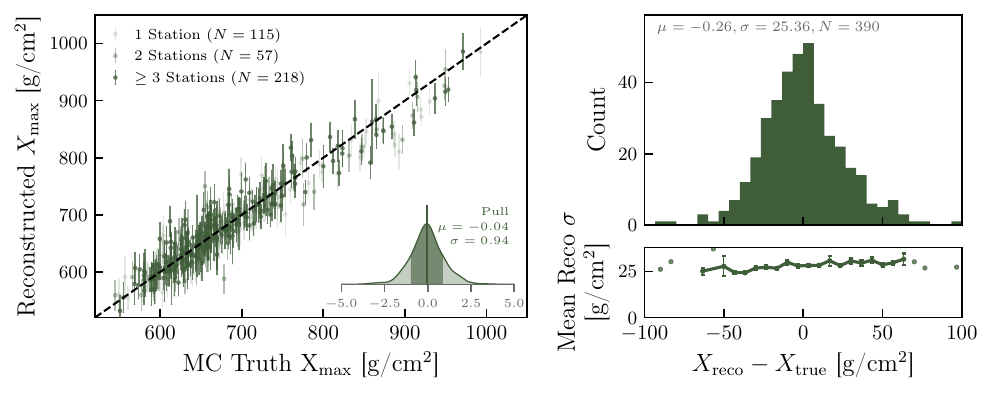}
    \caption{Performance of the $X_\text{max}$ reconstruction on CoREAS simulations. (Left) Reconstructed vs. true $X_\text{max}$ and pull distribution. The color-coding of the scatter points is based on how many stations were considered "triggered" for this specific event. (Right) Error distribution and uncertainty correlation.}
    \label{fig:xmax_sim}
\end{figure*}

Analogously, the performance on the $X_\text{max}$ is shown in Fig. \ref{fig:xmax_sim}. 
We achieve an overall precision on the $X_\text{max}$ of \SI{25.4}{g/cm^2} with a bias of \SI{-0.3}{g/cm^2}. The Pull $\sigma$ post-calibration of uncertainties is 0.94. Before calibration, the Pull $\sigma$ was 1.52. 

We observe that especially for larger shower maxima, where the footprints tend to extend only over $<\SI{100}{m}$, the reconstruction predicts lower shower maxima. Since both the fluence and the timing parameterization are less accurate for large $X_\text{max}$, this behaviour is somewhat expected.

Additionally to the combined timing-fluence model, we also tested an only fluence-based reconstruction. This was to test how much the timing information contributes to the fit. For this, we prepared the data exactly as for the timing-fluence case, but discarded timing data and the time component of the forward model. With that approach, we achieved a resolution in $X_\text{max}$ of \SI{36}{\gram\per\centi\meter\squared}, meaning the timing information improved the resolution substantially, as further discussed in Appendix \ref{app:timing}.

\subsubsection{Shower geometry performance}
The shower geometry being part of the forward model means the zenith angle, azimuth angle and core position are optimized during the reconstruction and we can access their posteriors. We find a precision on the zenith angle $\theta$ of \SI{0.5}{\degree} and an azimuth angle ($\phi$) precision of \SI{1}{\degree}, in line with the size of the LOFAR core and observation wavelength. The overall core accuracy achieved was \SI{8.4}{m}. 

The fluence-based fit, in which timing data was discarded, achieved a similar resolution in azimuth as the combined fit, but only a \SI{1.5}{\degree} resolution on the zenith angle. This further shows how the timing data is beneficial to the overall reconstruction (see Appendix \ref{app:timing}). 

\subsection{Impact of the trigger criterion}

\begin{figure}
    \centering
    \includegraphics[width=0.95\linewidth]{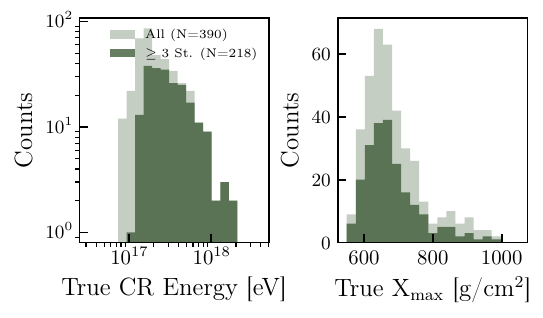}

    \caption{Distribution of primary energies and depths of shower maximum, showcasing the effect of loosening the trigger criterion.}
    \label{fig:distributions_e_xmax}
\end{figure}

Prior reconstructions placed a rather strict constraint on events that would be considered for a mass composition. In this work we have also considered events where just one station was triggered. This will increase the number of reconstructable events, allowing a mass composition study with improved statistics and extending to lower energies. We have shown the events from our whole simulation set in Fig. \ref{fig:distributions_e_xmax}, and marked the events that contained at least 3 triggered stations, and thus would have been considered in past analyses. As expected, more lower energy events made the less strict condition.

In a mass composition analysis these events would nevertheless have to be checked against a composition bias event by event, however \cite{Corstanje:2021} indicated that the previous trigger criterion was rather conservative.

\subsection{Resolution and Error Distribution Analysis}

\begin{figure}
    \centering
    \includegraphics[width=0.95\linewidth]{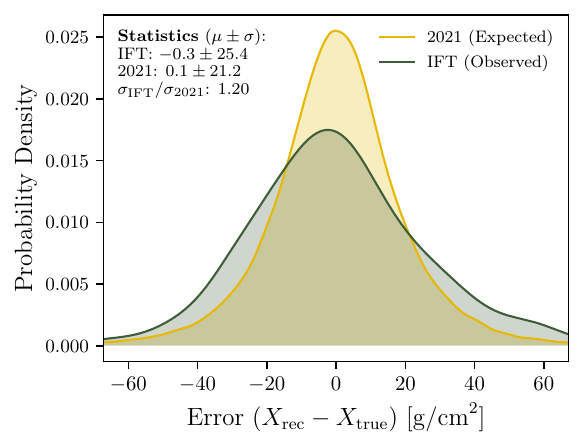}
    \caption{Comparison of the $X_\text{max}$ error distribution between the IFT method and the legacy 2021 method. The expected error from the legacy method is based on its distribution of reconstruction uncertainties, while the IFT method represents direct reconstruction errors. The ratio of the two is displayed in the Statistics legend.}
    \label{fig:errors_comparison}
\end{figure}

The resolution in $X_{\mathrm{max}}$ (Fig.~\ref{fig:errors_comparison}) as compared to the legacy method is evaluated by analyzing the distribution of signed errors, defined as the residual $\Delta X_{\mathrm{max}} = X_{\mathrm{rec}} - X_{\mathrm{true}}$. This contrasts the observed residuals from the IFT pipeline against the expected performance of the legacy method.

The IFT distribution (labeled `IFT') is derived directly from the calculated residuals of the filtered event set. The reference distribution (labeled `2021') is generated synthetically to model the theoretical resolution of the legacy pipeline, assuming unbiased reconstruction errors. This is achieved via a bootstrap resampling procedure:
\begin{enumerate}
    \item A pool of event uncertainties ($\sigma_{\mathrm{2021}}$) is extracted from the legacy dataset.
    \item For each event in the comparison set, a $\sigma_i$ is randomly sampled with replacement from this pool.
    \item A synthetic error is generated by drawing from a normal distribution $\mathcal{N}(\mu=0, \sigma=\sigma_i)$.
\end{enumerate}

Both distributions are visualized as continuous probability density functions using Gaussian Kernel Density Estimation (KDE). The arithmetic mean ($\mu$) and standard deviation ($\sigma$) for both distributions are calculated to quantify the bias and resolution, respectively, with the ratio $\sigma_{\mathrm{IFT}} / \sigma_{\mathrm{2021}}$ serving as a metric for relative precision. 

While the IFT method appears a factor $1.2$ less precise in this comparison, the resulting $\sim \SI{25}{g/cm^2}$ resolution remains better than the \SI{30}{}-\SI{40}{g/cm^2} typical of sparse arrays in this energy range \cite{Auger:2024xmax, Tunka-Rex:2015zsa} and approaches the benchmark of the legacy LOFAR analysis \cite{Corstanje:2021}. Furthermore, the achieved energy resolution of \SI{12.4}{\percent} competes well with the \SI{17}{\percent} precision reported by AERA \cite{PierreAuger:2016vya}, albeit this includes further detector effects. 

Combined with a computation time decrease of a factor of 1000 w.r.t. the legacy method and less strict triggering conditions allowing more "challenging" events (i.e. lower energies, smaller shower footprints) in our simulation set, this method is ideal for a mass composition study with a full LOFAR1.0 data set.

\section{Application to LOFAR data}
\begin{figure*}[t]
    \centering
    \includegraphics[width=0.8\linewidth]{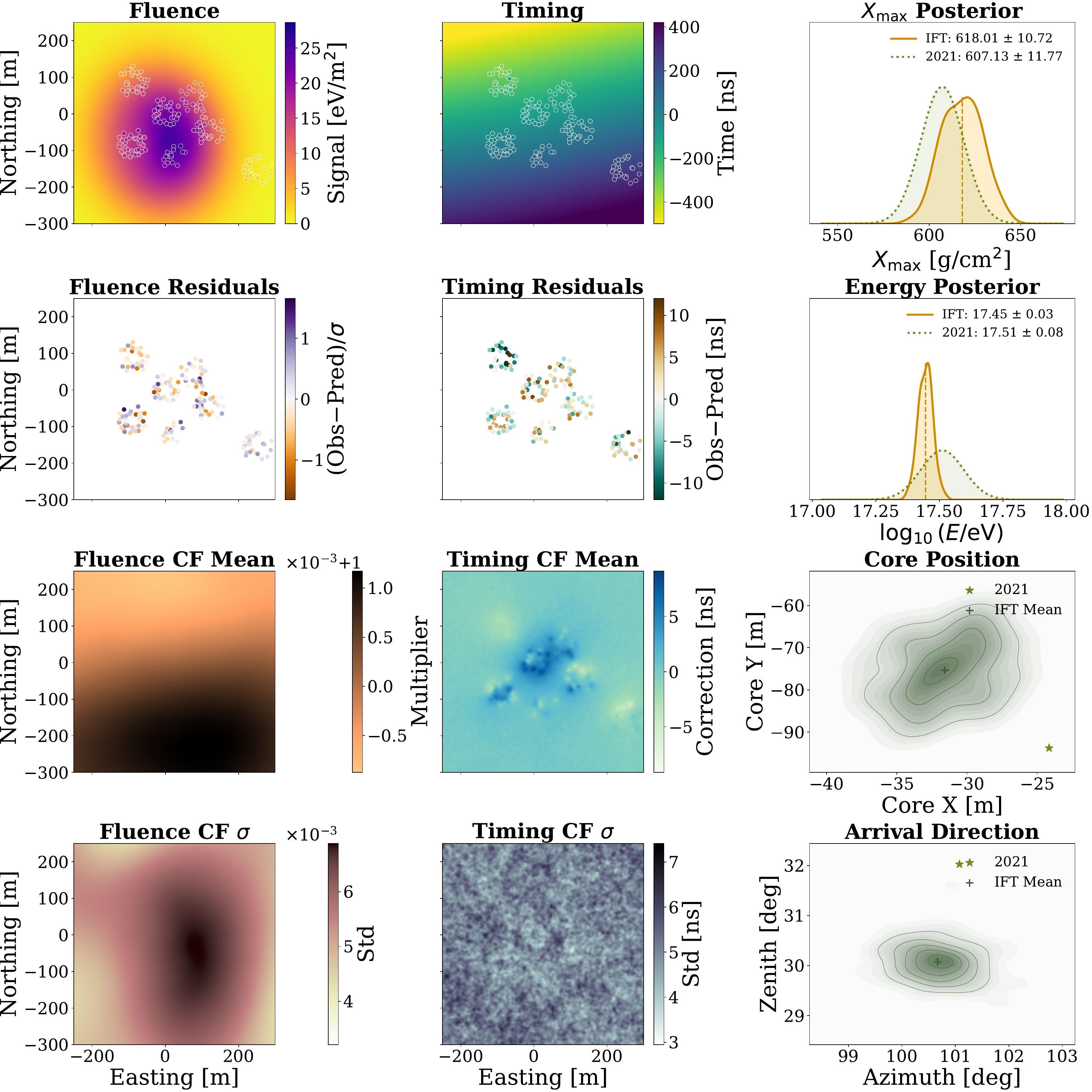}
    \caption{One of the LOFAR event reconstructions. Displayed are fluence in the left column, timing in the middle and posterior distributions compared with the legacy 2021 reconstruction on the, where the arrival direction was determined via a planewave fit. For the fluence and timing columns, the top row shows the data along with the posterior mean in the background. The second row shows the residuals and the third and fourth row the inferred Correlated Field mean and standard deviation.}
    \label{fig:reco_lofar_example_1}
\end{figure*}
To verify the method further, we select a subset of LOFAR events previously reconstructed in reference \cite{Corstanje:2021}. The data are read in and  converted into electric field traces using NuRadioReco \cite{GlaserNRR:2019,Bouma:2025dmo}.

We use the stored initial estimate of the LORA particle data fit on the core position and shower direction as our prior mean for the shower geometry, but maintain our own prior widths as used in Sec.~\ref{sec:hyperpriors}. The resulting prior distribution is usually wider than the expected LORA accuracy, but a wider prior can aid the optimizer in navigating through the posterior landscape and avoid getting stuck in local minima. This also makes us more independent of the goodness of fit of the particle data.

The pulse finding and data processing are handled as described in Sec.~\ref{sec:simtest}, without any further adjustment needed from what has been done for simulations. We use an individual atmosphere model for each event, matching the conditions of the event time.

Post-reconstruction we apply the simulation-based calibration described in Appendix \ref{app:uncertainty_calibration} to the reconstruction uncertainties to avoid underestimating them.

To fully compare the method with the legacy reconstruction, we need to calculate the primary energy $E_\text{CR}$ from the reconstructed radiation energy $E_\text{rad}^{30-\SI{80}{MHz}}$.  This can be done via the relation \cite{PierreAuger:2016vya}
\begin{equation}\label{eq:radiationtocr}
E_\text{rad}^{30-\SI{80}{MHz}}/\sin^2(\alpha)= A \times 10^7~\si{eV}(E_\text{CR}/10^{18}~\si{eV})^B.
\end{equation}
A more accurate relation would be only correcting the geomagnetic component of the radiation energy by $\sin^2(\alpha)$ \cite{Glaser:2016re}, however for the purpose of validating the method we use the simpler relation from Eq. \eqref{eq:radiationtocr}. We calibrate the parameters $A$ and $B$ on the reconstructed radiation energies of our simulation set (a mixed proton/iron composition), yielding $A=9.57\pm\SI{0.90}{MeV}$ and $B=2.010\pm0.011$. We note that this is smaller than the value previously obtained for LOFAR data \cite{Mulrey:2020oqe}, but consistent within systematic uncertainties. It still warrants further investigation when analyzing the whole LOFAR1.0 data-set, whether this is due to method choices like the time-integration window, the choice of hadronic interaction model, or a previously overlooked effect. For validating the method we have found this to be sufficiently self-consistent.

\subsection{Reconstruction quality}

Of the 289 LOFAR events, 231 reconstructions were successful, which is with about \SI{80}{\percent} similar to the simulations. For the unsuccessful reconstructions, the optimizer failed to achieve a reduced $\chi^2 < 1.25$, or returned a core uncertainty of > \SI{10}{m}. The test on simulations revealed that for those cases, the reconstruction accuracy was very low.

Post-reconstruction, we check for each of the samples generated during the optimization whether it would have triggered the required 3 stations. If not, the sample is discarded, ensuring that the posterior distribution only contains possible realizations of the model. In reality, only few ($<~\SI{1}{\percent}$) samples were discarded over the full event set, indicating that the optimizer rarely found unrealistic realizations.

An example reconstruction and its parameter comparison to the 2021 reconstruction is shown in Fig. \ref{fig:reco_lofar_example_1}. The residuals show little structure, indicating they are mostly noise and not signal. The agreement between the 2021 reconstruction and this work for $X_\text{max}$ and energy is good and in line with the behaviour discussed in Sec.~\ref{sec:e_rad_reco}. The legacy method only uses a planewave to describe the arrival direction, which explains the obtained offset in zenith angle.

\begin{figure}
    \centering
    \includegraphics[width=0.95\linewidth]{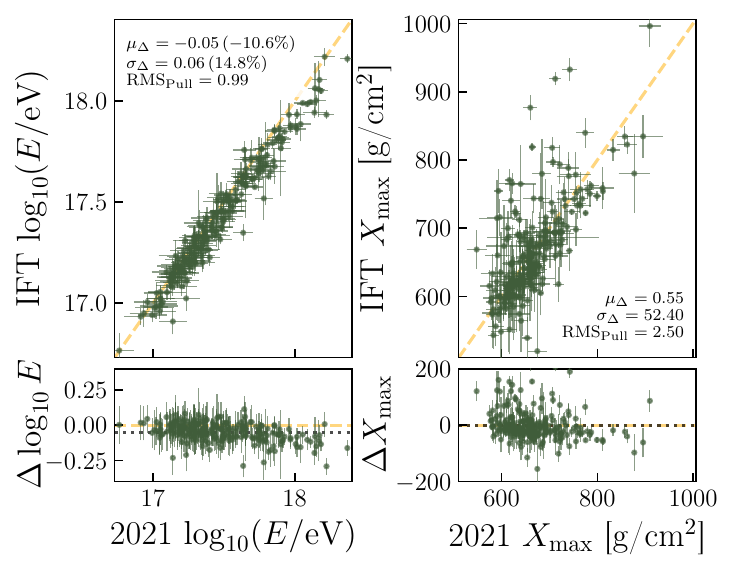}
    \caption{(Top) Comparison of reconstructed shower parameters $E_\text{CR}$ (left) and $X_\text{max}$ (right) by the legacy method (2021) and this work (IFT). The statistics in the top panel mark the mean, standard deviation and the standard deviation of the pull. The dashed black line in the bottom panel marks the mean bias.}
    \label{fig:event_wise_comparison}
\end{figure}

Furthermore, we show the event-wise comparison in reconstructed $X_\text{max}$ and $E_\text{CR}$ in Fig. \ref{fig:event_wise_comparison}. The $X_\text{max}$ distribution on the right shows no bias and a standard deviation of \SI{52.4}{\gram\per\centi\meter\squared}. Assuming the statistical uncertainties of the two methods are uncorrelated, the expected standard deviation of the difference distribution based on the respective resolutions would be $\sigma_\Delta=\sqrt{\sigma_\text{IFT}^2+\sigma_\text{2021}^2}\approx\SI{33}{\gram\per\centi\meter\squared}$. 
Upon further investigation of the differences in $X_\text{max}$ between the two methods, we notice that this large scatter is primarily outlier-driven, as the difference distribution is slightly asymmetric with a non-Gaussian tail, indicating a few strong outliers where the IFT method predicts much larger $X_\text{max}$ values than the legacy method. To account for this, we recalculated the scatter using a robust standard deviation estimator: the $68\%$ interquantile range, defined as $\sigma_{\mathrm{rob}} = (P_{84} - P_{16}) / 2$. This metric isolates the true $1\sigma$ width of the dense core of the data while ignoring the extreme outlier tails. Using this metric, we retrieve a scatter of only \SI{35}{\gram\per\centi\meter\squared}, however we also find that the median difference between the two methods is \SI{-9.7}{\gram\per\centi\meter\squared} using this metric. Nevertheless, for the absolute majority of reconstructions the methods agree within their nominal precisions for $X_\text{max}$. The strong outliers warrant further investigation for a future composition study. 

The reconstructed energies are systematically lower than in the legacy method, with the average reconstructed energy being \SI{10}{\percent} lower and with a standard deviation of \SI{15}{\percent}. As the energy bias seems to increase with energy, it cannot easily be corrected, however this might be due to the simplified conversion formula (Eq. \eqref{eq:radiationtocr}) that we used and will be investigated in the future before a full analysis of the LOFAR1.0 data set can be performed.

\subsection{Mass composition performance}

\begin{figure}
    \centering
    \includegraphics[width=0.95\linewidth]{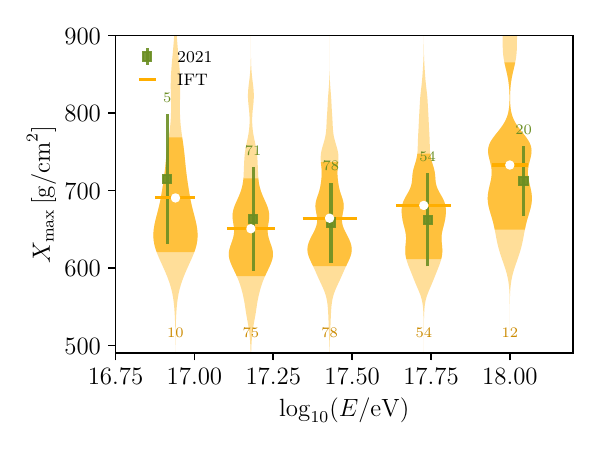}
    \caption{$X_\text{max}$ distributions as determined by the method (labeled "IFT"). The violin plot shows summed posterior distributions of the reconstructed $X_\text{max}$ for each energy bin, with the $1\sigma$ band highlighted. The horizontal error bar represents the statistical and systematic energy uncertainty. The legacy method (labeled "2021") is marked for comparison purposes. Error bars show both the statistical per-bin uncertainties and the event-by-event uncertainties. The numbers around each energy bin represent the number of events in the respective energy bin.}
    \label{fig:mass_composition}
\end{figure}

Figure \ref{fig:mass_composition} shows the $X_{\mathrm{max}}$ distributions of the reconstructed event sample. This was not cross-checked against composition biases and purely serves as a comparison between what the two methods predict for the same LOFAR event set.
To facilitate a direct comparison between the reconstruction pipelines, the dataset is strictly filtered to include only those events successfully reconstructed by both the legacy method and the new Information Field Theory approach. The data are grouped into energy bins of width $\Delta \log_{10}(E/\mathrm{eV}) = 0.3$ over the range $16.5 \leq \log_{10}(E/\mathrm{eV}) \leq 18.5$. The number of events contributing to each bin ($N$) is annotated explicitly in the figure.

For the legacy reconstruction, the binned data are represented by their arithmetic means and standard deviations. The square markers indicate the mean energy and mean depth of shower maximum ($X_{\mathrm{max}}$) of the events within the bin. The associated vertical and horizontal error bars represent the standard deviations of the $X_{\mathrm{max}}$ and $\log_{10}E$ distributions, respectively.

For the IFT reconstruction, the visualization leverages the full posterior probability distributions provided by the algorithm. Rather than reducing each event to a single scalar value, the posterior $X_{\mathrm{max}}$ samples from all events falling within a specific energy bin are summed. The resulting distribution is visualized using a Gaussian Kernel Density Estimation (KDE) to illustrate the total probability density of the reconstructed mass composition within that bin. The white circular markers indicate the global mean of these stacked samples, while the horizontal error bars denote the standard deviation of the mean reconstructed energies of the constituent events.

Within uncertainties, we observe good agreement between both methods.

\section{Summary and Outlook}

Our novel Bayesian approach using Information Field Theory has shown promising results with an accuracy comparable to the method previously used for LOFAR data. The main improvements are the factor of 1000 decrease in runtime w.r.t. the direct simulation comparison and the wider selection of events we were able to reconstruct in tests on simulations. The added timing information as well as the non-parametric additions to the model make the reconstruction significantly more accurate than a solely fluence-based fit, and more flexible in the application on real data. On realistic simulations, we achieved an $X_\text{max}$ resolution of \SI{25.4}{\gram\per\centi\meter\squared} and a resolution in $E_\text{rad}$ of \SI{12.4}{\percent}. The application to a small set of LOFAR data revealed no significant bias in $X_\text{max}$ and slight bias of \SI{10}{\percent} in primary energy.

In the future, further investigation of the impact of the fluence Correlated Field will be performed, both to find out what impact it has on shower parameters and also whether it is beneficial to be added to data. Overall, a deeper insight in the behaviour of the Correlated Fields on real data will be helpful, as they can model systematic behaviour and spatial correlations that a simple uncorrelated noise model cannot.

While tested on LOFAR data, we note that the method can in principle be used site-independently for near-vertical air shower reconstruction in the $30-\SI{80}{MHz}$ band. The implementation of the method along with the full forward model and nifty8 optimizer is available for use \cite{git:link}.

Future improvements of the method will entail reconstructing based on the full voltage traces, as they also contain information on key shower parameters. While parameterizations of the electric field spectra already exist \cite{Martinelli:2023}, these were developed for inclined showers and found to not hold up for near-vertical showers. A development of a parameterization for these spectra is thus necessary. Additionally, the LOFAR antennas specifically are highly resonant, which makes the reconstruction of a spectrum across the full bandwidth challenging. 

Future radio telescopes such as the SKA, which will have a more uniform antenna response over the full bandwidth, might be more suited for this approach. We are nevertheless working on the full waveform reconstruction, along with extensions of this model to broader frequency bands for LOFAR2.0 and the SKA.

\section*{Acknowledgments}

This work was supported by the German Federal Ministry for Research, Technology, and Space as part of the projects ErUM-IFT and D-LOFAR-ERIC, as well as the German Research Foundation through project 531213488. 

We also acknowledge support by the Dutch Research Council (NWO) grant number OCENW.XS25.1.237.

\appendix
\begin{figure*}
    \centering
    \includegraphics[width=0.95\linewidth]{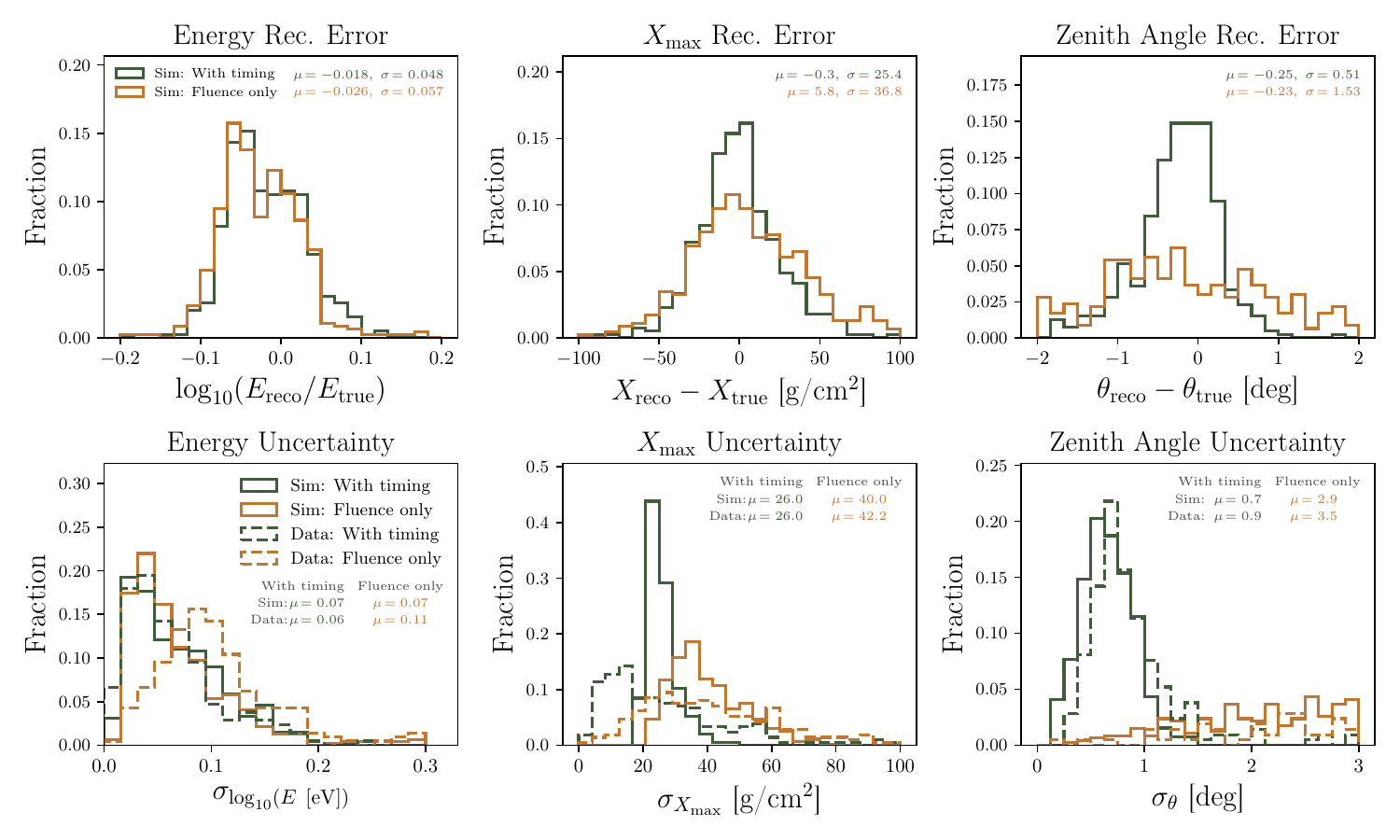}
    \caption{Resolution comparison of the pure fluence forward model compared to the model with timing information for three shower parameters (radiation energy, $X_\text{max}$ and zenith angle). The top panel shows the error distributions of reconstructed value compared to the simulated MC truth, with the statistics of each in the top right corner of each respective subfigure. The bottom panel shows uncertainty distributions both for simulations (solid lines) and data (dashed lines) derived from the IFT reconstruction, comparing fluence-only (yellow) and the full model (green). The mean uncertainty values are displayed in a table in the top right corner of each subfigure.}
    \label{fig:fluenceOnly}
\end{figure*}

\begin{figure}
    \centering
    \includegraphics[width=0.99\linewidth]{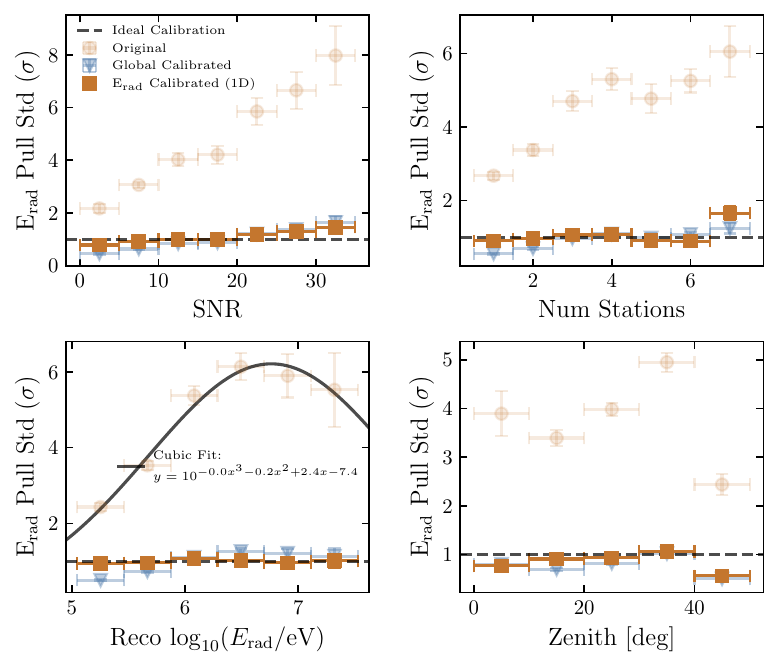}
    \caption{Pull standard deviation of the $E_\text{rad}$ for different shower parameters: SNR (maximum fluence SNR across the event), number of triggered stations, reconstructed radiation energy and zenith angle. The original Pull $\sigma$ are compared to a simple global correction and a correction based on a cubic fit to the Pull $\sigma$ as function of reconstructed radiation energy.}
    \label{fig:cal_erad}
\end{figure}

\begin{figure}
    \centering
    \includegraphics[width=0.99\linewidth]{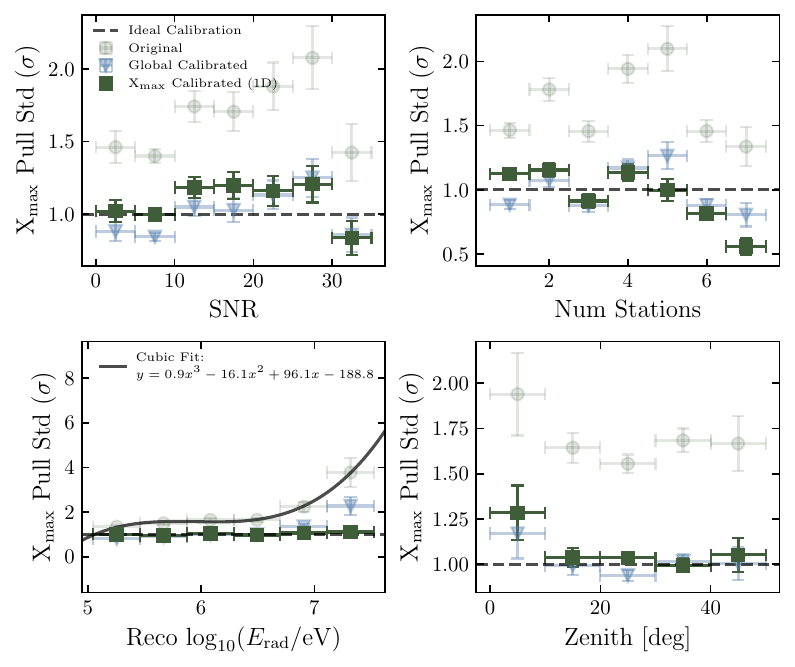}
    \caption{Pull standard deviation of the $X_\text{max}$ for different shower parameters: SNR (maximum fluence SNR across the event), number of triggered stations, reconstructed radiation energy and zenith angle. The original Pull $\sigma$ are compared to a simple global correction and a correction based on a cubic fit to the Pull $\sigma$ as function of reconstructed radiation energy.}
    \label{fig:cal_xmax}
\end{figure}

\label{app:timing}
\section{Impact of added timing information}

To determine the benefit of adding timing to the forward model, we performed a separate test on simulation and data using only the fluence implementation of the forward model. We kept all other steps identical to the data processing described in  Sec.~\ref{sec:simtest}, including prior widths, pulse finding and noise modeling. For the simulation set we can compare the resolution with respect to the MC truth, which is displayed in Fig.~\ref{fig:fluenceOnly} (top panel) for the radiation energy, the $X_\text{max}$ and the zenith angle.

The figure not only shows a clear improvement of resolution in $X_\text{max}$ and zenith angle when using the full forward model with timing for simulations, but it also shows that the estimated uncertainties are lower both for data and also for simulations, to a similar extent, as compared to a solely fluence-based reconstruction (bottom panel).
\label{app:timing}

\section{Calibration of uncertainties}\label{app:uncertainty_calibration}

To account for the underestimation of uncertainties, we calibrate based on the Pull $\sigma$ of our simulation set. Fig.~\ref{fig:cal_erad} shows the Pull $\sigma$ as a function of shower parameters for the radiation energy. 

The reconstruction appears overly confident for higher energy events, likely because for lower energies the data are more dominated by noise than by systematic differences from the forward model. We model the Pull $\sigma$ as a function of reconstructed radiation energy via a third degree polynomial fit. The posterior distribution of the radiation energy is then scaled by this function. Fig.~\ref{fig:cal_erad} also shows how a global correction by the overall Pull $\sigma$ would perform. Overall we find the $E_\text{rad}^\text{reco}$-based correction yields a better coverage.

The $X_\text{max}$ reconstruction in Fig.~\ref{fig:xmax_sim} also indicates undercoverage. As for the radiation energy, we find that the Pull increases as a function of energy, so we calibrate based on reconstructed radiation energy. Fig.~\ref{fig:cal_xmax} shows the Pull dependence for the $X_\text{max}$ as function of shower parameters, the cubic fit used for the calibration and the comparison with a simple global correction.

We note that there seems to be an additional dependence on zenith angle, as the bottom right subfigure of Fig.~\ref{fig:cal_xmax} indicates. We attempted also a two-dimensional calibration based on both radiation energy and zenith angle, however we found that the resulting overall coverage in $X_\text{max}$ was worse than for a one-dimensional calibration.

We suspect that, especially for the radiation energy reconstruction, the usage of correlated fields in the fluence forward model would provide a better coverage due to the added degrees of freedom in the model. In the future, the effect of the fluence correlated field on $X_\text{max}$ could be further investigated, possibly resulting in a similar reconstruction precision and better coverage without the need for a simulation-based calibration.

\section{Particle Model}
\label{app:ParticleModel}
The NKG function describes the particle density $\rho(r)$ at lateral distance $r$ from the shower axis as:
\begin{equation}
    \rho(r) = N_e \cdot C(s) \cdot \left( \frac{r}{R_M} \right)^{s-2} \left( 1 + \frac{r}{R_M} \right)^{s-4.5},
\end{equation}
where $N_e$ is the total number of particles at ground level, $s$ is the shower age parameter, and $R_M$ is the Molière radius. The normalization constant $C(s)$ ensures proper integration over the lateral plane:
\begin{equation}
    C(s) = \frac{\Gamma(4.5-s)}{2\pi R_M^2 \, \Gamma(s) \, \Gamma(4.5-2s)}.
\end{equation}

The shower age $s$ characterizes the evolutionary stage of the cascade and is typically $s \approx 1$ at shower maximum. For showers observed at ground level, $s > 1$ indicates post-maximum development. The total particle number $N_e$ is related to the primary cosmic ray energy $E_\text{CR}$ via an empirical relation:
\begin{equation}
\log_{10}(N_e^\text{max}) = \frac{\log_{10}(E_\text{CR}/\text{eV}) - A}{B},
\end{equation}
where $A$ and $B$ are constants calibrated to simulations.

An attenuation factor accounts for the decrease in particle density beyond shower maximum. This is modeled as:
\begin{equation}
\ln(N_e^\text{ground}) = \ln(N_e^\text{max}) - \frac{D_{X_\text{max}}}{\Lambda_\text{att}},
\end{equation}
where $\Lambda_\text{att}$ is the attenuation length for electromagnetic particles in air.

The LORA detectors do not count particles directly but instead measure the total deposited energy.
This quantity depends not only on the total number of particles hitting the detector but also on the track length through the scintillator, which is approximately $L/\cos(\theta)$ for fully contained tracks, with $L$ representing the thickness of the scintillator.
In practice, the effective length is somewhat shorter because some particles leave or enter through the sides; nevertheless, this geometric effect largely compensates for the reduction in the projected detector area.
Consequently, the predicted count in detector $i$ with effective area $A_{\text{eff}}$ is:
\begin{equation}
    \lambda_i = \rho(r_i) \cdot A_{\text{eff}} + n_\text{bg},
\end{equation}
where $n_\text{bg}$ represents a small background term.

As with the radio model, the particle prediction includes a non-parametric correlated field to model systematic uncertainties. This field is clipped to prevent unphysical corrections. The full implementation is differentiable, sharing the core position and arrival direction with the radio model while maintaining separate particle-specific parameters ($s$, $R_M$, overall scale). In the future one could even link the particle-specific parameters to the radio parameters, as the shower age depends on $X_\text{max}$ and the Molière radius can be parameterized from energy, $X_\text{max}$, arrival direction and the atmospheric model. This however goes beyond the scope of this work as for now we do not actively use the particle model for reconstruction.

\bibliographystyle{model1-num-names}

\bibliography{references}

@article{Kampert:2012,
title = {{Measurements of the cosmic ray composition with air shower experiments}},
journal = {Astroparticle Physics},
volume = {35},
number = {10},
pages = {660-678},
year = {2012},
issn = {0927-6505},
doi = {https://doi.org/10.1016/j.astropartphys.2012.02.004},
url = {https://www.sciencedirect.com/science/article/pii/S0927650512000382},
author = "Kampert, K.H. and Unger, M."
}

@article{Huege:2016,
title = {{Radio detection of cosmic ray air showers in the digital era}},
journal = {Physics Reports},
volume = {620},
pages = {1-52},
year = {2016},
issn = {0370-1573},
doi = {https://doi.org/10.1016/j.physrep.2016.02.001},
url = {https://www.sciencedirect.com/science/article/pii/S0370157316000636},
author = {Huege, T.}
}

@article{vanHaarlem:2013,
    author = "van Haarlem, M. P. and others",
    collaboration = "LOFAR",
    title = "{{LOFAR: The LOw-Frequency ARray}}",
    journal = "Astron. Astrophys.",
    volume = "556",
    pages = "A2",
    year = "2013"
}

@article{Mulrey:2025,
  author = "Mulrey, K. and others",
  title = "{{Cosmic-ray detection with LOFAR 2.0}}",
  doi = "10.22323/1.501.0340",
  journal = "PoS",
  year = 2025,
  volume = "ICRC2025",
  pages = "340"
}

@article{Thoudam:2014,
title = {{LORA: A scintillator array for LOFAR to measure extensive air showers}},
journal = {Nuclear Instruments and Methods in Physics Research Section A: Accelerators, Spectrometers, Detectors and Associated Equipment},
volume = {767},
pages = {339-346},
year = {2014},
issn = {0168-9002},
doi = {https://doi.org/10.1016/j.nima.2014.08.021},
url = {https://www.sciencedirect.com/science/article/pii/S0168900214009413},
author = {S. Thoudam and others},
}

@article{Huege:2013,
    author = {Huege, T. and Ludwig, M. and James, C. W.},
    title = {{Simulating radio emission from air showers with CoREAS}},
    journal = {AIP Conference Proceedings},
    volume = {1535},
    number = {1},
    pages = {128-132},
    year = {2013},
    issn = {0094-243X},
    doi = {10.1063/1.4807534},
    url = {https://doi.org/10.1063/1.4807534},
    eprint = {https://pubs.aip.org/aip/acp/article-pdf/1535/1/128/11832917/128_1_online.pdf},
}

@article{Mulrey:2019,
title = "{{Calibration of the LOFAR low-band antennas using the Galaxy and a model of the signal chain}}",
collaboration = {LOFAR},
journal = {Astroparticle Physics},
volume = {111},
pages = {1-11},
year = {2019},
issn = {0927-6505},
doi = {https://doi.org/10.1016/j.astropartphys.2019.03.004},
url = {https://www.sciencedirect.com/science/article/pii/S0927650518302810},
author = {Mulrey, K. and others},
}

@article{ensslin2019,
  author = {En{\ss}lin, Torsten A.},
  title = {{Information field theory}},
  journal = {Annalen der Physik},
  volume = {531},
  pages = {1800127},
  year = {2019},
  doi = {10.1002/andp.201800127}
}

@article{jasche2010,
  author = {Jasche, Jens and others},
  title = {{Bayesian power-spectrum inference for large-scale structure data}},
  journal = {Monthly Notices of the Royal Astronomical Society},
  volume = {406},
  pages = {60},
  year = {2010},
  doi = {10.1111/j.1365-2966.2010.16610.x}
}

@article{edenhofer2024,
  author = {Edenhofer, Gordian and others},
  title = "{{A parsec-scale Galactic 3D dust map out to 1.25 kpc from the Sun}}",
  journal = {Astronomy \& Astrophysics},
  volume = {685},
  pages = {A82},
  year = {2024},
  doi = {10.1051/0004-6361/202347628}
}

@article{welling2021,
  author = {Welling, Christoph and others},
  title = {{Reconstructing non-repeating radio pulses with Information Field Theory}},
  journal = {Journal of Cosmology and Astroparticle Physics},
  volume = {2021},
  number = {04},
  pages = {071},
  year = {2021},
  doi = {10.1088/1475-7516/2021/04/071}
}

@misc{strahnz2024,
  author = {Str{\"a}hnz, Simon and others},
  title = "{{Electric Field Reconstruction with Information Field Theory}}",
  year = {2024},
  howpublished = {PoS(ARENA2024)056},
  doi = {10.22323/1.470.0056}
}

@inproceedings{terveer2025,
  author = {Terveer, K. and others},
  title = "{{Reconstructing air showers using probabilistic forward modelling for LOFAR}}",
  booktitle = {39th International Cosmic Ray Conference},
  journal = {PoS(ICRC2025)},
  pages = {413},
  year = {2025},
  doi = {10.22323/1.501.0413}
}

@inproceedings{watanabe2025,
  author = {Watanabe, K. and others},
  title = "{{A novel approach for air shower profile reconstruction with dense radio antenna arrays using Information Field Theory}}",
  booktitle = {39th International Cosmic Ray Conference},
  journal = {PoS(ICRC2025)},
  pages = {436},
  year = {2025},
  doi = {10.22323/1.501.0436}
}

@article{knollmuller2020,
  author = {Knollm{\"u}ller, Jakob and En{\ss}lin, Torsten A.},
  title = "{{Metric Gaussian Variational Inference}}",
  year = {2020},
  eprint = {1901.11033},
  archivePrefix = {arXiv},
  primaryClass = {stat.ML}
}

@article{frank2021,
  author = {Frank, Philipp and Leike, Reimar and En{\ss}lin, Torsten A.},
  title = {{Geometric Variational Inference}},
  journal = {Entropy},
  volume = {23},
  number = {7},
  pages = {853},
  year = {2021},
  doi = {10.3390/e23070853}
}

@article{edenhofer2024niftyre,
  author = {Edenhofer, Gordian and others},
  title = "{{Re-Envisioning Numerical Information Field Theory (NIFTy.re): A Library for Gaussian Processes and Variational Inference}}",
  journal = {Journal of Open Source Software},
  volume = {9},
  number = {98},
  pages = {6593},
  year = {2024},
  doi = {10.21105/joss.06593}
}

@misc{arras2019,
  author = {Arras, Philipp and others},
  title = "{{NIFTy 5: Numerical Information Field Theory}}",
  year = {2019},
  howpublished = {Astrophysics Source Code Library},
  eprint = {ascl:1903.008}
}

@article{steininger2019,
  author = {Steininger, Theo and others},
  title = "{{NIFTy 3 - Numerical Information Field Theory: A Python framework for Bayesian signal inference on HPC clusters}}",
  journal = {Annalen der Physik},
  volume = {531},
  pages = {1800290},
  year = {2019},
  doi = {10.1002/andp.201800290}
}

@article{selig2013,
  author = {Selig, Marco and others},
  title = "{{NIFTy: Numerical Information Field Theory. A versatile Python library for signal inference}}",
  journal = {Astronomy \& Astrophysics},
  volume = {554},
  pages = {A26},
  year = {2013},
  doi = {10.1051/0004-6361/201321236}
}

@misc{bradbury2018,
  author = {Bradbury, James and others},
  title = "{JAX: composable transformations of Python+NumPy programs}",
  year = {2018},
  howpublished = {\url{http://github.com/google/jax}}
}

@article{Ensslin:2009,
    author = "Enßlin, T. A. and Frommert, M. and Kitaura, F. S.",
    title = "{Information field theory for cosmological perturbation reconstruction and non-linear signal analysis}",
    journal = "Phys. Rev. D",
    volume = "80",
    pages = "105005",
    year = "2009"
}

@article{Glaser:2019,
    author = "Glaser, C. and others",
    title = "{An analytic description of the radio emission of air showers based on its emission mechanisms}",
    journal = "Astropart. Phys.",
    volume = "104",
    pages = "64--77",
    year = "2019"
}

@article{Askaryan:1962,
    author = "Askaryan, G. A.",
    title = "{Excess negative charge of an electron-photon shower and its coherent radio emission}",
    journal = "Sov. Phys. JETP",
    volume = "14",
    pages = "441",
    year = "1962"
}

@article{Mitra:2020mza,
    author = "Mitra, P. and others",
    title = "{Reconstructing air shower parameters with LOFAR using event specific GDAS atmosphere}",
    eprint = "2006.02228",
    archivePrefix = "arXiv",
    primaryClass = "astro-ph.HE",
    doi = "10.1016/j.astropartphys.2020.102470",
    journal = "Astropart. Phys.",
    volume = "123",
    pages = "102470",
    year = "2020"
}

@article{GlaserNRR:2019,
   title="{NuRadioReco: a reconstruction framework for radio neutrino detectors}",
   volume={79},
   ISSN={1434-6052},
   url={http://dx.doi.org/10.1140/epjc/s10052-019-6971-5},
   DOI={10.1140/epjc/s10052-019-6971-5},
   number={6},
   journal={The European Physical Journal C},
   publisher={Springer Science and Business Media LLC},
   author={Glaser, C. and others},
   year={2019},
   month=jun }

@article{Corstanje:2023,
   title={A high-precision interpolation method for pulsed radio signals from cosmic-ray air showers},
   volume={18},
   ISSN={1748-0221},
   url={http://dx.doi.org/10.1088/1748-0221/18/09/P09005},
   DOI={10.1088/1748-0221/18/09/p09005},
   number={09},
   journal={Journal of Instrumentation},
   publisher={IOP Publishing},
   author={Corstanje, A. and others},
   year={2023},
   month=sep, pages={P09005} }

@article{Corstanje:2021,
   title="{Depth of shower maximum and mass composition of cosmic rays from 50 PeV to 2 EeV measured with the LOFAR radio telescope}",
    collaboration = {LOFAR},
   volume={103},
   ISSN={2470-0029},
   url={http://dx.doi.org/10.1103/PhysRevD.103.102006},
   DOI={10.1103/physrevd.103.102006},
   number={10},
   journal={Physical Review D},
   publisher={American Physical Society (APS)},
   author={Corstanje, A. and others},
   year={2021},
   month=may }

@article{Martinelli:2023,
  author = "Martinelli, Sara  and  Schlüter, Felix  and  Huege, Tim",
  title = "{Parameterization of the frequency spectrum of radio emission in the 30-80 MHz band from inclined air showers}",
  doi = "10.22323/1.424.0036",
  journal = "PoS",
  year = 2023,
  volume = "ARENA2022",
  pages = "036"
}

@article{PierreAuger:2025hoe,
    author = "Halim, A. Abdul and others",
    collaboration = "Pierre Auger",
    title = "{Long-term calibration and validation of stability of the Auger Engineering Radio Array using the diffuse Galactic radio emission}",
    eprint = "2512.03692",
    archivePrefix = "arXiv",
    primaryClass = "astro-ph.IM",
    reportNumber = "FERMILAB-PUB-25-0895",
    doi = "10.1088/1748-0221/20/12/P12017",
    journal = "JINST",
    volume = "20",
    number = "12",
    pages = "P12017",
    year = "2025"
}

@article{Schellart:2013bba,
    author = "Schellart, P. and others",
    collaboration = {LOFAR},
    title = "{Detecting cosmic rays with the LOFAR radio telescope}",
    eprint = "1311.1399",
    archivePrefix = "arXiv",
    primaryClass = "astro-ph.IM",
    doi = "10.1051/0004-6361/201322683",
    journal = "Astron. Astrophys.",
    volume = "560",
    pages = "A98",
    year = "2013"
}

@article{Corstanje:2025wbc,
    author = "Corstanje, A. and others",
    title = "{LOFAR-style reconstruction of cosmic-ray air showers with SKA-Low}",
    eprint = "2504.16873",
    archivePrefix = "arXiv",
    primaryClass = "astro-ph.HE",
    doi = "10.1103/l8mt-994v",
    journal = "Phys. Rev. D",
    volume = "112",
    number = "2",
    pages = "023017",
    year = "2025"
}

@article{PierreAuger:2023lkx,
    author = "Abdul Halim, A. and others",
    collaboration = "Pierre Auger",
    title = "{Demonstrating Agreement between Radio and Fluorescence Measurements of the Depth of Maximum of Extensive Air Showers at the Pierre Auger Observatory}",
    eprint = "2310.19963",
    archivePrefix = "arXiv",
    primaryClass = "astro-ph.HE",
    reportNumber = "FERMILAB-PUB-24-0138-AD-CSAID-PPD-TD-V",
    doi = "10.1103/PhysRevLett.132.021001",
    journal = "Phys. Rev. Lett.",
    volume = "132",
    number = "2",
    pages = "021001",
    year = "2024"
}

@article{PierreAuger:2016vya,
    author = "Aab, Alexander and others",
    collaboration = "Pierre Auger",
    title = "{Measurement of the Radiation Energy in the Radio Signal of Extensive Air Showers as a Universal Estimator of Cosmic-Ray Energy}",
    eprint = "1605.02564",
    archivePrefix = "arXiv",
    primaryClass = "astro-ph.HE",
    reportNumber = "FERMILAB-PUB-16-169-AD-AE-CD-TD",
    doi = "10.1103/PhysRevLett.116.241101",
    journal = "Phys. Rev. Lett.",
    volume = "116",
    number = "24",
    pages = "241101",
    year = "2016"
}

@article{AbdulHalim:2025xz,
  author = "Abdul Halim, Adila  and  others",
collaboration = {Pierre Auger},
  title = "{Determination of the energy scale of cosmic ray measurements using the Auger Engineering Radio Array}",
  doi = "10.22323/1.501.0292",
  journal = "PoS",
  year = 2025,
  volume = "ICRC2025",
  pages = "292"
}

@article{Mulrey:2020oqe,
    author = "Mulrey, K. and others",
    collaboration = {LOFAR},
    title = "{On the cosmic-ray energy scale of the LOFAR radio telescope}",
    eprint = "2005.13441",
    archivePrefix = "arXiv",
    primaryClass = "astro-ph.HE",
    doi = "10.1088/1475-7516/2020/11/017",
    journal = "JCAP",
    volume = "11",
    pages = "017",
    year = "2020"
}

@article{Buitink:2016nkf,
    author = "Buitink, S. and others",
    collaboration = {LOFAR},
    title = "{A large light-mass component of cosmic rays at $10^{17}$ - $10^{17.5}$ eV from radio observations}",
    eprint = "1603.01594",
    archivePrefix = "arXiv",
    primaryClass = "astro-ph.HE",
    doi = "10.1038/nature16976",
    journal = "Nature",
    volume = "531",
    pages = "70",
    year = "2016"
}

@article{Tunka-Rex:2015zsa,
    author = "Bezyazeekov, P. A. and others",
    collaboration = "Tunka-Rex",
    title = "{Radio measurements of the energy and the depth of the shower maximum of cosmic-ray air showers by Tunka-Rex}",
    eprint = "1509.05652",
    archivePrefix = "arXiv",
    primaryClass = "hep-ex",
    doi = "10.1088/1475-7516/2016/01/052",
    journal = "JCAP",
    volume = "01",
    pages = "052",
    year = "2016"
}

@article{Tunka-Rex:2016nto,
    author = "Apel, W. D. and others",
    collaboration = "Tunka-Rex, LOPES",
    title = "{A comparison of the cosmic-ray energy scales of Tunka-133 and KASCADE-Grande via their radio extensions Tunka-Rex and LOPES}",
    eprint = "1610.08343",
    archivePrefix = "arXiv",
    primaryClass = "astro-ph.IM",
    doi = "10.1016/j.physletb.2016.10.031",
    journal = "Phys. Lett. B",
    volume = "763",
    pages = "179--185",
    year = "2016"
}

@article{Schellart:2014oaa,
    author = "Schellart, P. and others",
    collaboration = {LOFAR},
    title = "{Polarized radio emission from extensive air showers measured with LOFAR}",
    eprint = "1406.1355",
    archivePrefix = "arXiv",
    primaryClass = "astro-ph.HE",
    doi = "10.1088/1475-7516/2014/10/014",
    journal = "JCAP",
    volume = "10",
    pages = "014",
    year = "2014"
}

@article{deVries:2011pa,
    author = "de Vries, K. D. and van den Berg, A. M. and Scholten, O. and Werner, K.",
    title = "{Coherent Cherenkov Radiation from Cosmic-Ray-Induced Air Showers}",
    eprint = "1107.0665",
    archivePrefix = "arXiv",
    primaryClass = "astro-ph.HE",
    doi = "10.1103/PhysRevLett.107.061101",
    journal = "Phys. Rev. Lett.",
    volume = "107",
    pages = "061101",
    year = "2011"
}

@article{Allan:1972wd,
    author = "Allan, H. R.",
    title = "{Low frequency radio emission from extensive air showers}",
    doi = "10.1038/237384a0",
    journal = "Nature",
    volume = "237",
    pages = "384--385",
    year = "1972"
}

@article{James:2010vm,
    author = "James, Clancy W. and Falcke, Heino and Huege, Tim and Ludwig, Marianne",
    title = "{General description of electromagnetic radiation processes based on instantaneous charge acceleration in `endpoints'}",
    eprint = "1007.4146",
    archivePrefix = "arXiv",
    primaryClass = "physics.class-ph",
    doi = "10.1103/PhysRevE.84.056602",
    journal = "Phys. Rev. E",
    volume = "84",
    pages = "056602",
    year = "2011"
}

@article{Nelles:2014dja,
    author = "Nelles, A. and others",
    collaboration = {LOFAR},
    title = "{Measuring a Cherenkov ring in the radio emission from air showers at 110{\textendash}190 MHz with LOFAR}",
    eprint = "1411.6865",
    archivePrefix = "arXiv",
    primaryClass = "astro-ph.IM",
    doi = "10.1016/j.astropartphys.2014.11.006",
    journal = "Astropart. Phys.",
    volume = "65",
    pages = "11--21",
    year = "2015"
}

@article{Buitink:2014eqa,
    author = "Buitink, S. and others",
    collaboration = {LOFAR},
    title = "{Method for high precision reconstruction of air shower $X_{max}$ using two-dimensional radio intensity profiles}",
    eprint = "1408.7001",
    archivePrefix = "arXiv",
    primaryClass = "astro-ph.IM",
    doi = "10.1103/PhysRevD.90.082003",
    journal = "Phys. Rev. D",
    volume = "90",
    number = "8",
    pages = "082003",
    year = "2014"
}

@article{Corstanje:2014waa,
    author = "Corstanje, A. and others",
    collaboration = {LOFAR},
    title = "{The shape of the radio wavefront of extensive air showers as measured with LOFAR}",
    eprint = "1404.3907",
    archivePrefix = "arXiv",
    primaryClass = "astro-ph.HE",
    doi = "10.1016/j.astropartphys.2014.06.001",
    journal = "Astropart. Phys.",
    volume = "61",
    pages = "22--31",
    year = "2015"
}

@article{Greisen:1960,
    author = "Greisen, K.",
    title = "{Cosmic Ray Showers}",
    doi = "10.1146/annurev.ns.10.120160.000431",
    journal = "Ann. Rev. Nucl. Sci.",
    volume = "10",
    pages = "63--108",
    year = "1960"
}

@article{Kamata:1958,
    author = "Kamata, K. and Nishimura, J.",
    title = "{The lateral and the angular structure functions of electron showers}",
    doi = "10.1143/PTPS.6.93",
    journal = "Prog. Theor. Phys. Suppl.",
    volume = "6",
    pages = "93--155",
    year = "1958"
}

@article{Apel:2014,
doi = {10.1088/1475-7516/2014/09/025},
collaboration = {LOPES},
url = {https://doi.org/10.1088/1475-7516/2014/09/025},
year = {2014},
month = {sep},
publisher = {},
volume = {2014},
number = {09},
pages = {025},
author = {Apel, W.D. and others},
title = {The wavefront of the radio signal emitted by cosmic ray air showers},
journal = {Journal of Cosmology and Astroparticle Physics},
}

@article{Bouma:2025dmo,
    author = {Bouma, Sjoerd and Desmet, Mitja and Glaser, Christian and Laub, Philipp and Nelles, Anna and Pyras, Lilly and Schl{\"u}ter, Felix and Terveer, Karen},
    title = "{Integrating radio detectors of cosmic-ray air showers into the open-source NuRadio framework}",
    doi = "10.22323/1.501.0345",
    journal = "PoS",
    volume = "ICRC2025",
    pages = "345",
    year = "2025"
}

@article{Auger:2024xmax,
  title = "{Radio measurements of the depth of air-shower maximum at the Pierre Auger Observatory}",
  author = {Abdul Halim, A. and others},
  collaboration = {Pierre Auger Collaboration},
  journal = {Phys. Rev. D},
  volume = {109},
  issue = {2},
  pages = {022002},
  numpages = {24},
  year = {2024},
  month = {Jan},
  publisher = {American Physical Society},
  doi = {10.1103/PhysRevD.109.022002},
  url = {https://link.aps.org/doi/10.1103/PhysRevD.109.022002}
}

@article{Glaser:2016re,
doi = {10.1088/1475-7516/2016/09/024},
url = {https://doi.org/10.1088/1475-7516/2016/09/024},
year = {2016},
month = {sep},
publisher = {},
volume = {2016},
number = {09},
pages = {024},
author = {Glaser, Christian and Erdmann, Martin and Hörandel, Jörg R. and Huege, Tim and Schulz, Johannes},
title = {Simulation of radiation energy release in air showers},
journal = {Journal of Cosmology and Astroparticle Physics},
}

@article{PierreAuger:2015nim,
title = "{The Pierre Auger Cosmic Ray Observatory}",
journal = {Nuclear Instruments and Methods in Physics Research Section A: Accelerators, Spectrometers, Detectors and Associated Equipment},
author = {Aab, A. and others},
collaboration = {Pierre Auger Collaboration},
volume = {798},
pages = {172-213},
year = {2015},
issn = {0168-9002},
doi = {https://doi.org/10.1016/j.nima.2015.06.058},
url = {https://www.sciencedirect.com/science/article/pii/S0168900215008086},
}

@article{DESMET2026,
title = "{SMIET: Fast and accurate synthesis of radio pulses from extensive air shower using simulated templates}",
journal = {Astroparticle Physics},
volume = {175},
pages = {103182},
year = {2026},
issn = {0927-6505},
doi = {https://doi.org/10.1016/j.astropartphys.2025.103182},
url = {https://www.sciencedirect.com/science/article/pii/S0927650525001057},
author = {Mitja Desmet and Stijn Buitink and Tim Huege and Keito Watanabe}
}

@MASTERSTHESIS{Cantarini:2025,
      author       = {Cantarini, Giulia},
      othercontributors = {Nelles, Anna and Huesca Santiago, Enrique and Terranova,
                          Francesco},
      title        = {{M}odeling the reflection of radio signalsfrom atmospheric
                      clouds as a near-horizonbackground for astroparticle
                      observatories},
      school       = {University of Milano Bicocca},
      type         = {Masterarbeit},
      reportid     = {PUBDB-2026-00436},
      pages        = {92},
      year         = {2025},
      cin          = {Z-RAD},
      cid          = {I:(DE-H253)Z-RAD-20210408},
      pnm          = {613 - Matter and Radiation from the Universe (POF4-613)},
      pid          = {G:(DE-HGF)POF4-613},
      experiment   = {EXP:(DE-H253)RNO-G-20230101},
      typ          = {PUB:(DE-HGF)19},
      url          = {https://bib-pubdb1.desy.de/record/644602},
}

@article{Arras:2022,
title = "{Variable structures in M87* from space, time and frequency resolved interferometry}",
journal = {Nature Astronomy},
volume = {6},
pages = {259-269},
year = {2022},
doi = {https://doi.org/10.1038/s41550-021-01548-0},
url = {https://www.nature.com/articles/s41550-021-01548-0},
author = {Arras, Philipp and others}
}

@article{strahnz2025,
  author = "Strähnz, Simon  and  Huege, Tim  and  Enßlin, Torsten  and  Terveer, Karen  and  Nelles, Anna",
  title = "{Information Field Theory based Event Reconstruction for Cosmic Ray Radio Detectors}",
  doi = "10.22323/1.501.0402",
  journal = "PoS",
  year = 2025,
  volume = "ICRC2025",
  pages = "402"
}

@article{Huege2008,
doi = {10.1088/1742-6596/110/6/062012},
url = {https://doi.org/10.1088/1742-6596/110/6/062012},
year = {2008},
month = {may},
publisher = {},
volume = {110},
number = {6},
pages = {062012},
author = {Huege, T. and others},
collaboration = {LOPES},
title = {Radio detection of cosmic ray air showers with the LOPES experiment},
journal = {Journal of Physics: Conference Series},
}

@misc{Ravn2025,
      title={Likelihood Reconstruction for Radio Detectors of Neutrinos and Cosmic Rays}, 
      author={Martin Ravn and Christian Glaser and Thorsten Glüsenkamp and Ayca Öcelikkale and Alan Coleman},
      year={2025},
      eprint={2510.21925},
      archivePrefix={arXiv},
      primaryClass={astro-ph.IM},
      url={https://arxiv.org/abs/2510.21925}, 
}

@article{Ostapchenko_2013,
  author    = {Ostapchenko, S.},
  title     = {{QGSJET-II}: physics, recent improvements, and results for air showers},
  journal   = {EPJ Web of Conferences},
  volume    = {52},
  year      = {2013},
  pages     = {02001},
  month     = {jun},
  doi       = {10.1051/epjconf/20135202001},
  url       = {https://doi.org/10.1051/epjconf/20135202001},
  note      = {ISVHECRI 2012 -- XVII International Symposium on Very High Energy Cosmic Ray Interactions}
}

@misc{git:link,
  author = {Terveer, Karen},
  title = "{Air shower reconstruction code}",
  year = {2026},
  howpublished = {\url{https://github.com/karenterveer/RIA}}
}

\end{document}